\documentclass[a4paper]{emulateapj}
\usepackage{hyperref}

\shorttitle{Direct collapse in metal-enriched galaxy mergers}
\shortauthors{Mayer et al.}
\slugcomment{Submitted to ApJ}

\begin{document}

\title{Direct formation of supermassive black holes in metal-enriched gas at the heart of high-redshift galaxy mergers}

\author{Lucio Mayer\altaffilmark{1}, Davide Fiacconi\altaffilmark{1}, Silvia Bonoli\altaffilmark{2},
Thomas Quinn\altaffilmark{3}, Rok Roskar\altaffilmark{4}, Sijing Shen\altaffilmark{5}, James Wadsley\altaffilmark{6}}
\altaffiltext{1}{Center for Theoretical Astrophysics and Cosmology, Institute for Computational Science, \& Physik Institut, University of Zurich, Winterthurerstrasse 190, CH-8057 Z\"{u}rich, Switzerland}
\altaffiltext{2} {Centro de Estudios de Fisica del Cosmos de Aragon, Plaza San Juan 1, Planta-2, 44001, Teruel, Spain}
\altaffiltext{3}{Astronomy Department, University of Washington, Box 351580, Seattle, WA, 98195-1580, USA}
\altaffiltext{4}{Scientific IT Services, ETH Z\"{u}rich, Weinbergstrasse 11, CH-8092 Z\"{u}rich, Switzerland}
\altaffiltext{5}{Institute of Astronomy, University of Cambridge, Madingley Road, Cambridge CB3 0HA, UK}
\altaffiltext{6}{Department of Physics and Astronomy, McMaster University, Hamilton, Ontario L8S 4M1, Canada}


\begin{abstract}
We present novel 3D multi-scale SPH simulations of gas-rich galaxy mergers between the most massive 
galaxies at $z \sim 8 - 10$, designed to scrutinize the direct collapse formation scenario for massive black hole seeds proposed in \citet{mayer+10}.
The simulations achieve a resolution of 0.1 pc, and include both metallicity-dependent optically-thin cooling and a model for thermal balance at high optical depth.
We consider different formulations of the SPH hydrodynamical equations, including thermal and metal diffusion.
When the two merging galaxy cores collide, gas infall produces a compact, optically thick nuclear disk 
with densities exceeding $10^{-10}$ g~cm$^3$. 
The disk rapidly accretes higher angular momentum gas from its surroundings reaching $\sim 5$ pc and a mass of $\gtrsim 10^9$ $M_{\odot}$ in only a few $10^4$ yr.
Outside $\gtrsim 2$ pc it fragments into massive clumps.
Instead, supersonic turbulence prevents fragmentation in the inner parsec region, which remains warm ($\sim 3000-6000$ K) and develops strong non-axisymmetric modes
that cause prominent radial gas inflows ($> 10^4$ $M_{\odot}$~yr$^{-1}$), forming an ultra-dense massive disky core.
Angular momentum transport by non-axisymmetric modes should continue below our spatial resolution limit, quickly turning the disky  core into a supermassive protostar which can
collapse directly into a massive black hole of mass $10^8-10^9$ $M_{\odot}$ via the relativistic radial instability. 
Such a ``cold direct collapse'' explains naturally the early emergence of high-z QSOs. Its telltale signature would be a burst of gravitational waves 
in the frequency range $10^{-4} - 10^{-1}$ Hz, possibly detectable by the planned {\it eLISA} interferometer. 
\end{abstract}



\section{Introduction}

The origin of supermassive black holes (SMBHs) is still a puzzle. 
At least three different classes of models have been proposed: light seeds from Pop III stars, massive seeds from direct gas collapse and light to intermediate mass seeds formed inside star clusters via runaway mergers of stars and stellar remnants.
Light seeds from Pop III stars originate at $z > 12-15$ and are the natural consequence of primordial star formation \citep{madau+01, volonteri+03,tanaka+09}.
However, they may face difficulties in growing at a fast enough rate, at least comparable to the Eddington rate, to produce black holes exceeding $10^9$ $M_{\odot}$ in less than half a billion years \citep{johnson+07,pelupessy+07,wise+08,alvarez+09,milosavljevic+09}, 
as required by the existence of high-$z$ quasars (QSOs; \citealt{fan+01,mortlock+11, treister+13}).
Pop III seeds indeed tend to grow at rates well below Eddington as they 
are surrounded by low density ionized gas and are hosted in small halos that cannot sustain fast inflows of fresh gas from the cosmic web due to their shallow potential wells.
Seeds from dense star clusters form when the cluster is subject to rapid mass segregation and consequent runaway mergers that lead to the formation of a central very massive star $\gtrsim 1000$ $M_{\odot}$ (VMS; \citealt{gurkan+04,freitag+06}).
The VMS is then expected to end its life leaving behind a BH seed with $\sim 100-1000$ $M_{\odot}$ \citep[see e.g.][]{portegies-zwart+04,belkus+07}.
However, intense stellar wind may strongly suppress the growth of the VMS \citep[e.g.][]{vink+08,glebbeek+09}, unless it evolves in low metallicity environments such as high-$z$ 
nuclear star cluster \citep{devecchi+09,devecchi+12}.
The resulting BH seed might be more massive if the entire star cluster is pushed towards rapid collapse by mass loading
from a large scale gas inflow \citep{davies+11,lupi+14}.
Finally, direct collapse models rely on gravitational instabilities, such as spiral instabilities and bars-in-bars, to transport gas all the way from galactic disk scales 
to sub-pc scales \citep{begelman+06,lodato+06,begelman+09}. As a result, a Jeans unstable cloud forms and it may later evolve into a supermassive 
star (SMS), which in turn produces a quasi-star (QS) nurturing a massive BH seed ($M \sim 10^4 -10^5$ $M_{\odot}$) at its center \citep{begelman+08,volonteri+10,dotan+11}.
Alternatively, if the cloud or supermassive star  is massive enough ($M > 10^5 - 10^6$ $M_{\odot}$), and has negligible rotational support, it can collapse via the general relativistic
radial instability all the way into a black hole encompassing more than half of its original mass \citep{hoyle+63, baumgarte+99,shibata+02,saijo+09,montero+12}.
However, conventional direct collapse scenarios posit suppression of
radiative cooling in order to avoid fragmentation and form a single supermassive object \citep{lodato+06,dijkstra+06}. This in turn 
requires primordial gas composition and efficient dissociation of one of the most efficient primordial coolants, namely molecular hydrogen.
These requirements can be fulfilled under special environmental conditions such as strong external Lyman-Werner ionizing flux \citep[see][]{ferrara+13a,dijkstra+14}.
The conditions arising in direct collapse models may also restart the growth of a dormant Pop III seed that has been previously accreting inefficiently \citep{begelman+12,madau+14}.
Interestingly, cosmological simulations have shown convincingly that cold gas accretion flows originating from the cosmic web can sustain a mass accretion rate high enough to produce
the high-$z$ QSOs on the required short timescale in rare peaks at $z > 6$ (halos with virial mass $M_{\rm vir} \sim 10^{12} - 10^{13}$ $M_{\odot}$), provided that BH seeds in such halos
start with a mass $> 10^5$ $M_{\odot}$ and that they accrete constantly near the Eddington rate
\citep{dimatteo+12}.

An alternative scenario for direct collapse has been proposed by \citet[][hereafter MA10]{mayer+10}, and later tested 
against observational constraints of $z>6$ QSOs using a semi-analytical model attached to the {\it Millennium Simulation} 
by \citet{bonoli+14}.
In such a scenario, massive BH seeds originate from the collapse of a supermassive cloud assembled at the end of $z \sim 4 -10$ gas-rich galaxy mergers by multi-scale gas inflows driven by gravoturbulence and gravitational torques. 
The model does not require primordial gas composition, rather it is designed to be at work in very massive galaxies that have already been enriched to solar metallicity at $z < 10$.
MA10 measure gas inflow rates $\dot{M} \sim 10^3-10^4$ $M_{\odot}$~yr$^{-1}$ at parsec scales and below.
The merger-driven model postulates that the formed seeds grow fast but are rare, since they require gas-rich, major mergers between highly biased host halos corresponding to $4-5\sigma$ fluctuation \citep{bonoli+14}.
On the small scale side, the model predicts inflows that well exceed those necessary to build a SMS \citep[see][and references therein]{bonoli+14}, while other direct collapse models based on instabilities in protogalaxies at $z > 10$ fall short of the required inflow rates \citep[e.g.][]{regan+09a,regan+09b,latif+12}.

The model, however, has been questioned since it is based on simulations with idealized thermodynamical conditions.
Indeed, the simulations of MA10 employed an effective equation of state (EoS) with a variable polytropic index across a wide range of gas densities, valid for metal-enriched gas.
It has been pointed out that the adopted EoS may artificially promote the formation of a central supermassive gas cloud by suppressing gas fragmentation and star formation.
Instead, the expected high cooling rates should lead to widespread fragmentation, disrupting the gas inflow and leading to a BH of only $\sim 100$ $M_{\odot}$ in the most optimistic scenario \citep{ferrara+13b}.
The latter arguments, however, are based on simple one-dimensional disk models evolved in the isochoric limit, which forces the
density to remain constant at a given radius, neglecting the effect of shocks. These models also neglect self-gravity.
Therefore they lack the self-consistent coupling between thermodynamics and gasdynamics.
It is conceivable that self-gravity is a crucial ingredient 
in the MA10 model since the inflow is generated by multi-scale 
gravitational torques and it is ultimately mediated by a massive, self-gravitating, marginally unstable nuclear gas disk at scales $\lesssim 100$ pc.

Here we present 3D hydrodynamic simulations with initial conditions identical to MA10 but with a much richer inventory of physical processes.
Most notably, we include radiative cooling in both the optically-thin and optically-thick regimes in place of a prescribed EoS. 
We even explore the effect of different formulations of SPH, including a novel implementation of the SPH hydro force along with thermal diffusion which successfully obviate to the notorious difficulties of capturing fluid instabilities and mixing (\citealt{keller+14} -- see also \citealt{ritchies+01} and \citealt{hopkins+13}).

As we show in this paper, the basic picture proposed in MA10 is confirmed and further strengthened.
Not only the inflow occurs and forms a supermassive cloud even under these more realistic thermodynamical conditions, but it is 
enhanced by cooling, producing a much more compact and denser cloud on a shorter timescale.
As a result, we are left with the intriguing possibility that the collapse proceeds all the way to direct SMBH formation via the relativistic radial instability.
Despite the metal-line cooling, fragmentation is limited due to shock heating and gravoturbulence.
We conclude that, contrary to previous claims, our scenario for direct collapse is tenable. It offers an explanation for the rapid emergence of high-$z$ QSOs without requiring any special conditions in the interstellar medium (ISM), such as primordial gas composition and dissociation of molecular hydrogen.

\begin{deluxetable*}{lcccc}
\tablecaption{List of performed simulations with their characteristics \label{table_sim}}
\tablehead{\colhead{Label} & \colhead{SPH force} & \colhead{cooling type} & \colhead{thermal diffusion} & \colhead{metal diffusion} \\}
\startdata
RUN1 &  standard & metal cooling & no & no \\
RUN2 & standard & metal cooling & yes & yes \\
RUN3 &  standard &  metal cooling + thermal balance model & yes & yes \\
RUN4 & GDSPH & metal cooling & yes & yes \\
\enddata
\end{deluxetable*}

The paper is organized as follows: in Section \ref{sec_2} we describe the setup of the simulations and the implementation of the physical processes included.
We present our findings in Section \ref{sec_3}, 
while we discuss possible evolution scenarios of our final configuration towards a massive BH seed in Section \ref{sec_evol}.
In Section \ref{sec_4} we highlight the limitations that could affect our results and we present our conclusions.


\section{Numerical Simulations} \label{sec_2}


\subsection{Initial conditions and particle splitting}

The main purpose of this work is to study the gas dynamics of a gas-rich major galaxy merger from kiloparsec scales down to $\sim 0.1$ pc.
To this aim, we start by running a large scale galaxy merger simulation and then we apply particle splitting in the later phases of the evolution to reach sub-pc resolution, following MA10.
Since we want to compare with our previous work, we adopt the same initial conditions of the reference large-scale merger simulation used in MA10, namely a prograde coplanar 1:1 merger.
The merger simulation was initially run with a mass resolution of $2 \times 10^4$ $M_{\odot}$ for the gas particles and a gravitational softening of 100 pc for all galaxy components.
The simulation includes radiative cooling, star formation and supernova feedback (see \citealt{kazantzidis+05,mayer+07}; MA10 for further details).
The galaxies approach on a typical parabolic orbit inferred from cosmological simulations \citep{khochfar+06}.
The significant computational burden demanded by these simulations favors the choice of a coplanar
prograde merger, which leads to full coalescence of the two galaxy cores on a timescale somewhat shorter than in the other configurations \citep{kazantzidis+05,capelo+14}.
Nonetheless,  the mass of gas that concentrates at scales $\gtrsim 100$ pc, which is a key feature in our scenario, has been shown to be rather insensitive to the merger geometry for a fixed mass ratio between the two galaxies \citep{kazantzidis+05}.

The two galaxy models represent identical multi-component systems with virial mass $M_{\rm vir} = 10^{12}$ $M_{\odot}$. 
They are constructed following the method by \citet{hernquist+93} and choosing the structural parameters in agreement with the scaling laws predicted by the $\Lambda$-CDM model \citep{mo+98, kazantzidis+05}.
The models include a dark matter halo that follows the Navarro-Frank-White \citep[NFW;][]{navarro+96} profile with concentration $c=12$, an exponential stellar and gaseous disk and a stellar bulge.
The stellar disk has mass $M_{\rm disk} = 6 \times 10^{10}$ $M_{\odot}$, scale radius $R_{\rm disk} = 3.5$ kpc and scale height $z_{\rm disk} = 350$ pc.
The gaseous disk has the same scale lengths of the stellar disk.
The mass is about $0.1 M_{\rm disk}$ at the time of the last pericenter passage, when star formation has already consumed part of the gas.
The bulge is a \citet{hernquist+90} model with mass $M_{\rm bulge} = M_{\rm disk}/5$ and scale length $r_{\rm bulge} = 0.7$ kpc.
We refer to MA10 for more details on the numerical parameters of the large scale merger simulation \citep[see also][]{kazantzidis+05, mayer+07}.

We note that the adopted virial mass is comparable to that of a Milky Way-sized galaxy at $z=0$, and it corresponds to $\sim 5 \sigma$ fluctuations at $z = 8-9$ in the {\it WMAP9} cosmology.
This is consistent with the very low abundance and high clustering amplitudes of high-$z$ QSOs \citep{volonteri+06,bonoli+14}.
We have chosen model parameters that are similar to those of a present-day, massive, disk-dominated galaxy, which is supported by recent cosmological simulations showing that typical disk dominated galaxies at $z >3$ have structural properties and morphologies already akin to their counterparts at low $z$, with a disk and a sizable bulge component \citep{fiacconi+14}.
However, gas fractions are often higher at high-$z$, in the range $0.3-0.6$, as suggested by both observations and simulations \citep{tacconi+10,moody+14,fiacconi+14}, and the most massive among high-$z$ galaxies might have hosted clumpy, turbulent disks \citep{tacconi+10,ceverino+12,forster+14,moody+14} rather than smooth exponential disks as in our initial conditions.
Therefore our initial conditions follow a conservative approach since a more gas-rich and gravoturbulent disk will promote loss of angular momentum and central inflow even before the merger occurs \citep[e.g.][]{bournaud+12}.

As in MA10,  we split individual gas particles within a spherical volume of radius  30 kpc into 8 child particles, when galaxy cores are 6 kpc apart at their last apocenter.
We reduce correspondingly their gravitational softening, achieving a resolution of $\sim 2600$ $M_{\odot}$ and 0.1 pc.
Pre-existing star particles and dark matter particles are not split, and their softening is left unvaried.
They essentially provide a smooth background potential to avoid spurious two-body heating against the much lighter gas particles.
The momentum conserving splitting procedure, which improves over that used in MA10, is described in \citet{roskar+14}. 
The choices of when to split the gas particles and the size of splitting volume are identical to that in MA10 and are discussed in the Supplementary Information of the latter paper.
They are motivated by inducing minimal numerical fluctuations  by introducing a refined region large enough to avoid any contamination of low-res particles for the entire duration
of the simulations. 
After the two cores reached a separation $< 100$ pc, most of the gaseous mass is collected in the inner 100 pc volume and it is traced with as many as 1.5 million gas particles.
By performing numerical tests we have verified that, owing to the fact that gas dominates the mass and dynamics of the nuclear region, the large softening adopted for the dark matter particles does not affect significantly the
density profile of the inner dark halo that surrounds the nuclear disk. These and other numerical tests showing the robustness of the technique were presented in \citet{mayer+07}.

\begin{figure*}
\epsscale{1.15}
\plotone{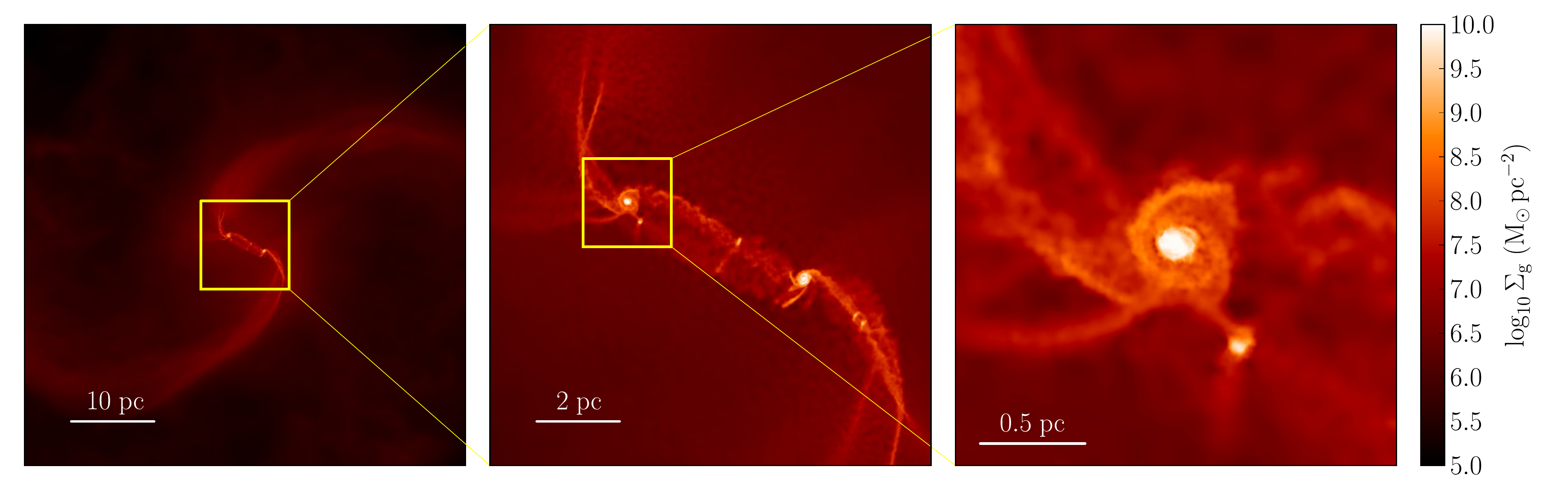}
\figcaption{Projected gas density of RUN4 taken just prior to the final merger of the two cores, at time $t = t_{0} - 3.2$ kyr, where $t_{0}$ is the time of coalescence of the approaching cores (see the text for details).
The zoom-in view to the inner few parsecs shows one of the two dense cores and a few clumps produced by fragmentation of the infalling gas.\label{fig_SD_init}}
\end{figure*}


\subsection{Features of the simulation suite} \label{sec_suite}

We have run four different versions of the second part of the simulation (i.e. after particle splitting) using GASOLINE2, a new version of the GASOLINE code \citep{wadsley+04, keller+14}.
Individual runs differ in the sub-grid model for radiative cooling as well as for the specific implementation of the SPH equations of hydrodynamics.
We describe below the different features of the runs and we summarize them in Table \ref{table_sim}.
In the following, we refer to each run with its own label according to Table \ref{table_sim}.
In all the runs we adopt the metal-dependent, optically-thin cooling introduced in \citet{shen+10}.
It considers tabulated cooling rates in ionization equilibrium, while for H and He we compute directly the rates without assuming equilibrium \citep{wadsley+04}.
As in MA10, we assume solar metallicity gas, consistent with observational constraints on the metallicity of the hosts of high-$z$ QSOs \citep{walter+04}.
In the following, we call this scheme ``metal cooling''.
In one of the runs (RUN3), the code switches to an equilibrium temperature-density relation above $\rho_{\rm g} = 0.1$ H~cm$^{-3}$.
This relation is calibrated on 2D radiative transfer calculations based on an improved version 
of the \citet{spaans+00} model, as described in detail in \citet{roskar+14}.
In the following, we call this additional feature ``thermal balance model''.
The ``thermal balance model'' includes: (i) metallicity-dependent opacity effects due to absorption and scattering of photons by dust; (ii) IR dust radiation; (iii) photoelectric effect on dust; (iv) atomic and molecular line trapping in an ISM irradiated by stellar light; and finally (v) heating by cosmic rays.
The thermal balance model thus accounts for self-shielding effects in the dense ISM.

The ``thermal balance model'' assumes the presence of a uniform UV photon radiation field produced by a starburst in gas of a specified metallicity.
Before the final merger, the large scale simulation show that a starburst with a strength of $\sim 100$ $M_{\odot}$~yr$^{-1}$ takes place in the inner kiloparsec (see MA10).
Therefore, we assume such a  star formation rate as input to determine the stellar UV flux required by the thermal balance model.  
While the intensity of the starburst is actually a free parameter, the assumed value is on the low side.
A starburst with ten times higher strength, which is likely at high-$z$, would produce a higher UV flux intensity, which in turn would heat up the dust and enhance photoelectric heating of the gas; this would produce a higher gas temperature floor and moderate net radiative losses, strengthening further the main findings of this paper concerning stability to fragmentation (see Section \ref{sec_3}).

\begin{figure*}
\epsscale{1.15}
\plotone{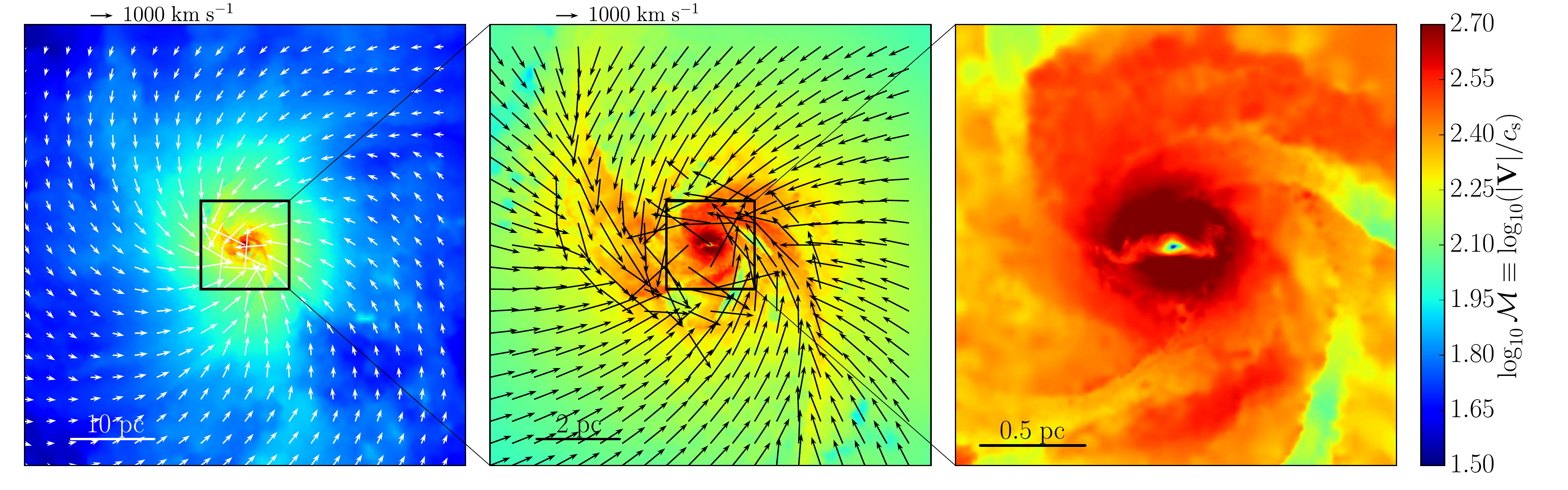}
\plotone{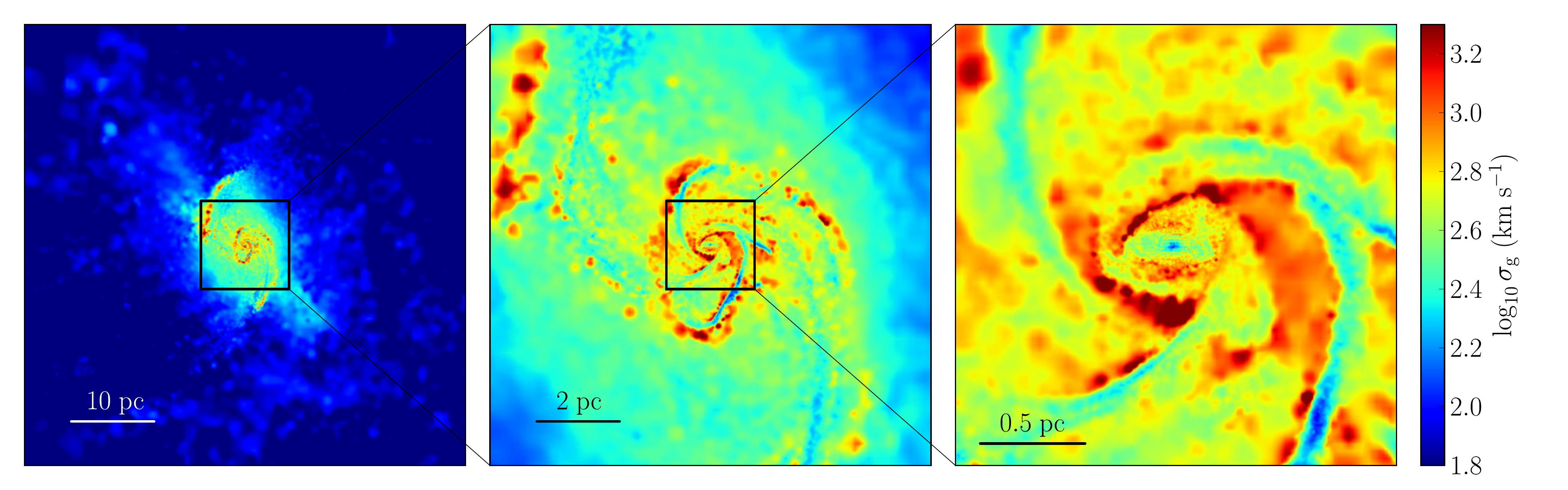}
\figcaption{Color-coded maps at different scales of the density-weighted Mach number $\mathcal{M} \equiv |\mathbf{V}|/c_{\rm s}$ (top row) and of the density-weighted 3D gas velocity dispersion (bottom row) at  time $t_0$ for RUN4.
The first two columns of the first row show the velocity field of the gas.
The flow is highly supersonic ($\mathcal{M} > 100$) and turbulent ($\sigma_{\rm g} \lesssim 1000$ km~s$^{-1}$ in the inner region), and it maintains this status till the end of the simulation.
\label{fig_mach}}
\end{figure*}
 
A pressure floor ensures that the Jeans mass is resolved locally by $\gtrsim 1$ resolution element, in order to avoid spurious fragmentation.
Following \citet{agertz+09}, the minimum pressure is set to $P_{\rm min} = \alpha G \max(h,\epsilon)^2 \rho^2$, where $\alpha= 3.0$ is a fudge factor, $G$ is the gravitational constant, $\rho$ is the density of the particle, and
$\max(h,\epsilon)$ is the local resolution element, i.e. the maximum between the smoothing length $h$ and the softening length $\epsilon$ of the gas particle.
All the simulations (including the the large-scale merger simulation before splitting is applied) adopt star formation  and blast-wave supernova feedback following \citet{stinson+06}.
In particular, we allow stars to form from gas following a Schmidt law in regions above the density threshold  
$n_{\rm th} = 10^4$ H~cm$^{-3}$ and below the temperature threshold $T_{\rm th} = 500$ K.

We also compare runs with different implementations of SPH that should capture hydrodynamic instabilities and multi-phase fluid mixing with increasing degree of realism.
In particular: (i) RUN1 employs the original SPH implementation of GASOLINE (dubbed in the following ``standard'' SPH; \citealt{wadsley+04}); (ii) RUN2 and RUN3 employ 
the ``standard'' SPH with the thermal and metal diffusion terms in the momentum and energy equations based on the sub-grid turbulence  prescription described by \citet{shen+10}; 
and (iii) RUN4 employs thermal and metal diffusion as well as the new implementation of the hydrodynamical force equation described by \citet{keller+14}.
The new approach is based on the geometric density average (GDSPH) in the SPH force expression $(P_i+P_j)/(\rho_i\rho_j)$ in place of the usual $P_i/\rho_i^2+P_j/\rho_j^2$, where $P_i$ and $\rho_i$ are particle pressures and densities respectively.
This modification leads to smoother gradients and removes artificial surface tension \citep{keller+14}.
Detailed tests of the GDSPH implementation combined with diffusion as in RUN4 will be presented in Wadsley et al. (in preparation), showing that it can successfully capture Kelvin-Helmoltz and Rayleigh-Taylor instabilities. 
Note that other new SPH implementations, while they often track entropy rather than energy as we do, also use a geometric density mean for the forces (e.g. \citealt{hopkins+13, saitoh+13}).

\begin{figure*}
\epsscale{1.15}
\plotone{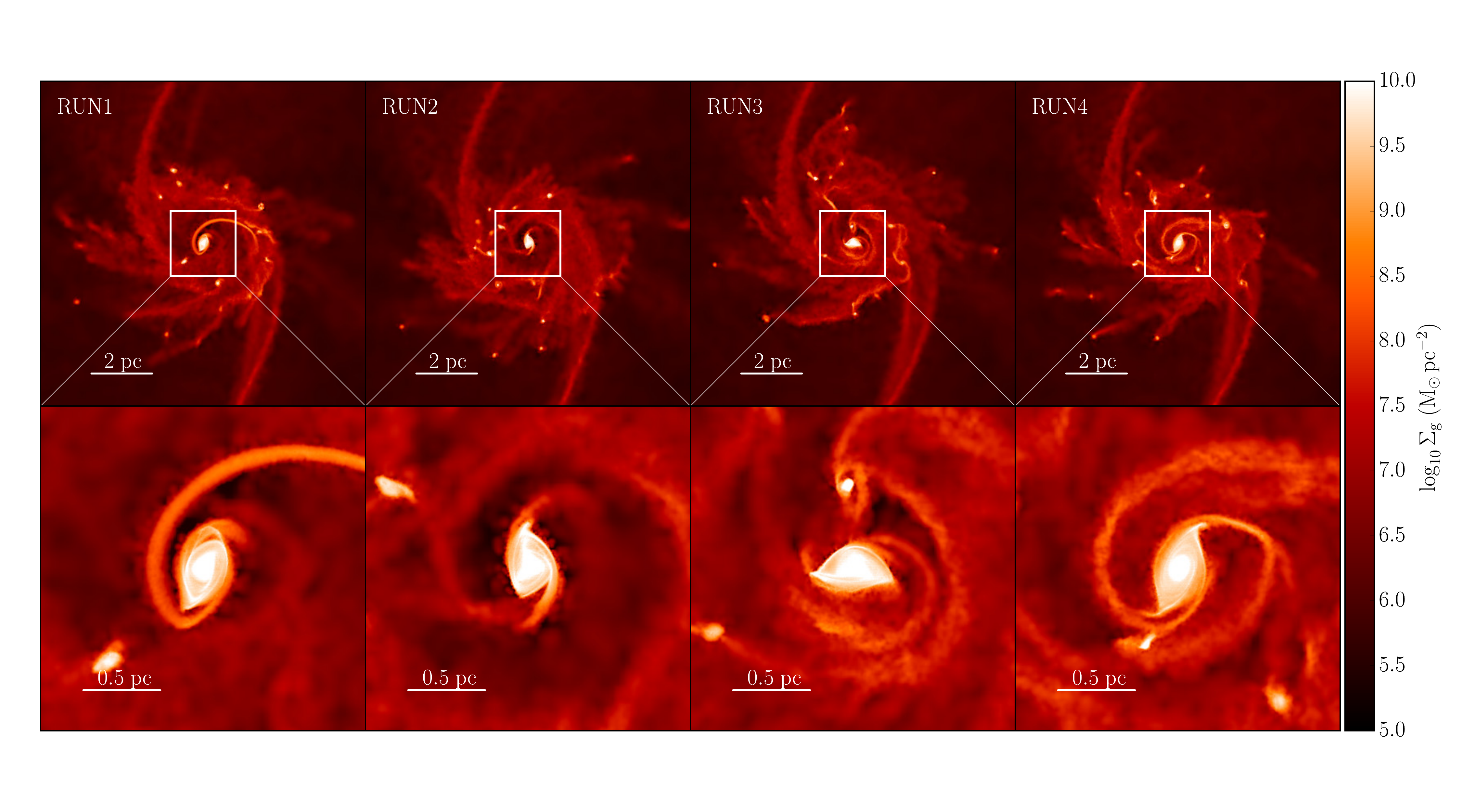}
\figcaption{Projected gas density maps of the nuclear region of the merger at $t_0 +  5$ kyr, showing a disk-like object with radius $\lesssim 5$ pc (upper row), and a 
compact inner disk-like core less than a pc in size (lower row). 
From left to right, the columns compare RUN1, RUN2, RUN3, and RUN4, respectively, with labels according to Table \ref{table_sim}. 
The larger scale disk shows some fragmentation, while the inner region is subject to strong non-axisymmetric instabilities, but negligible fragmentation.
The evolution is weakly sensible to the specific SPH implementation or to the adopted model for cooling.\label{fig_SD}}
\end{figure*}


\section{Results} \label{sec_3}


\subsection{Global evolution of the nuclear region}\label{sec_global}

We begin our analysis when the two galaxy cores are only $\sim 10$ pc apart and $\lesssim 10$ kyr away from final coalescence, as shown by the gas density projection in Figure \ref{fig_SD_init}.
As the two galaxy cores approach their last encounter, efficient cooling leads to some fragmentation within the colliding 
gas filaments surrounding them.
However, most of the gas flows inward unimpeded as the two cores sink owing to dynamical friction against the background matter.
Hereafter we define the reference merging time of the cores, $t_0$, as the time at which two separate density peaks cannot be identified any longer.  
Our analysis will focus on $t \geq t_0$.

The gas flow is radial, with velocities exceeding 1000 km~s$^{-1}$, and highly supersonic, as shown by the Mach number map at $t \geq t_0$ in the top row 
of Figure \ref{fig_mach} for RUN4.
We define the Mach number $\mathcal{M}$ as the ratio between the module of the gas 3D velocity $|\mathbf{V}|$ and the local thermal sound speed $c_{\rm s}$.
In the inner $\sim 5$ pc, $\mathcal{M} > 100$, leading to strong shocks and the development of supersonic turbulence, as shown by the 3D velocity dispersion maps in the bottom row of Figure \ref{fig_mach}.
The local velocity dispersion is measured as $\sigma_{\rm g} = \sqrt{\langle |\mathbf{V}|^2 \rangle - |\langle \mathbf{V} \rangle|^2}$, where $\mathbf{V}$ is the local 3D gas velocity and the average $\langle \cdot \rangle$ is intended as SPH average on the smoothing kernel.

After the final coalescence of the two cores, the residual angular momentum in the infalling gas establishes 
a centrifugal barrier and leads to the formation of a self-gravitating nuclear disk with radius $\sim 5$ pc, as shown in the gas density projections of Figure \ref{fig_SD}.
At the center, the gas accumulates in a compact, disky core that shows non-axisymmetric features such as $m=1-2$ modes.
This final configuration appears to be largely independent on the specific SPH implementation and cooling module adopted in the various runs (Figure \ref{fig_SD}) and is maintained till the end of the simulation.
We emphasize that the nuclear disk is extremely well resolved owing to particle splitting, with nearly half a million particles. 

\begin{figure}
\epsscale{1.1}
\plotone{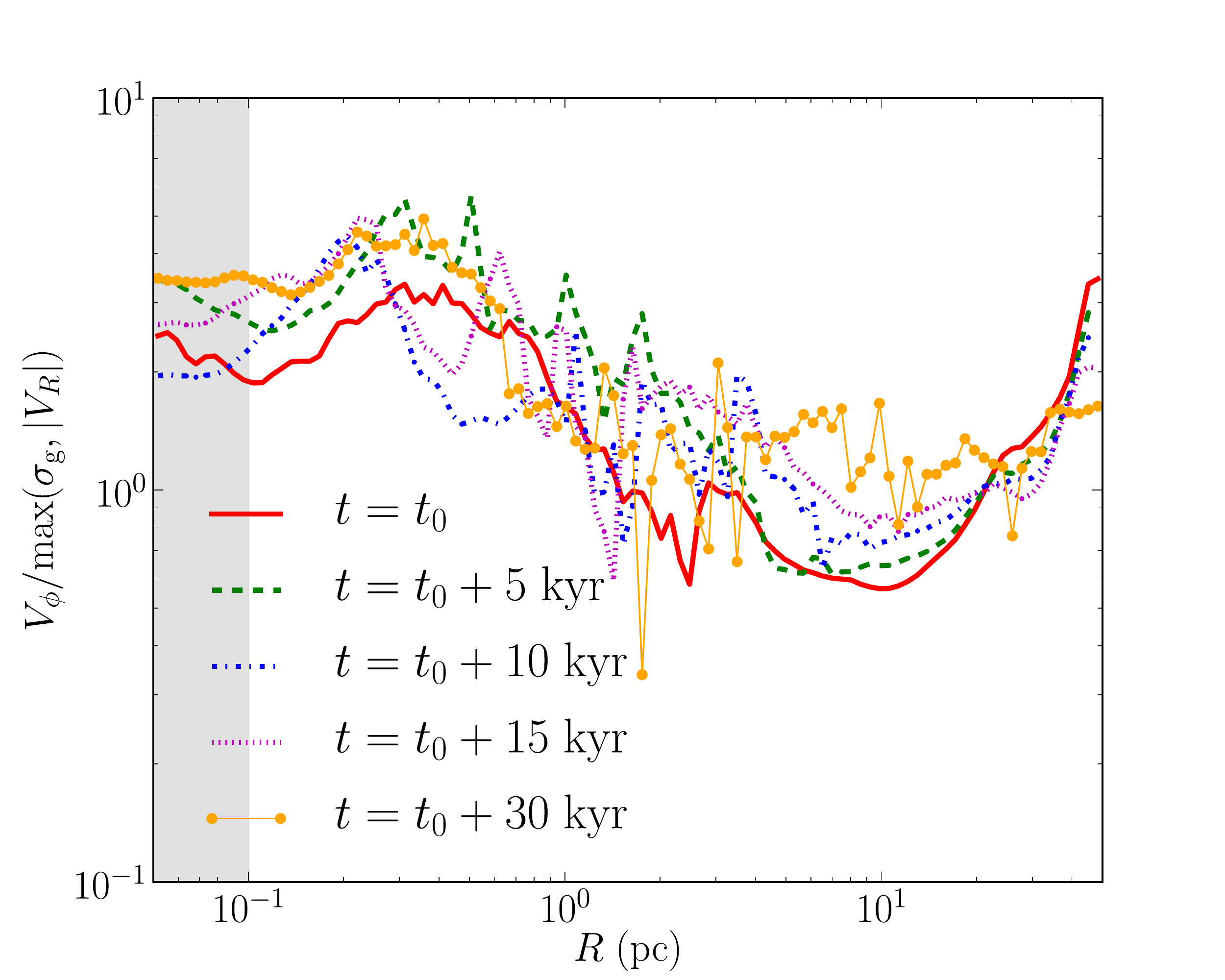}
\figcaption{Time evolution for RUN4 of the profile of the ratio between the azimuthal velocity $V_{\phi}$ and the maximum between the radial velocity $|V_{R}|$ and the local 3D velocity dispersion $\sigma_{\rm g}$.
The gray band highlights the resolution limit given by the gravitational softening.
\label{fig_velocity}}
\end{figure}

The central disk is surrounded by filamentary rings of accreting material which exhibit substantial rotation out to several pc during the whole simulation.
This is shown by the radial profile of $V_{\phi} / \max(\sigma_{\rm g}, |V_{R}|)$ in Figure \ref{fig_velocity}.
$V_{\phi} / \max(\sigma_{\rm g}, |V_{R}|)$ quantifies the balance between rotation and random motions by means of the ratio between the azimuthal velocity $V_{\phi}$ and the maximum between the radial velocity and the local velocity dispersion $\max(\sigma_{\rm g}, |V_{R}|)$.
In all runs the central disk undergoes only moderate fragmentation despite being massive and self-gravitating.
Indeed, the rotating disk and surrounding infalling filamentary rings are stable except along the densest spiral arms, where gravitationally bound clumps are produced via fragmentation (the stability of the nuclear region is described with detail in Section \ref{sec_stability}).

\begin{figure}
\epsscale{1.1}
\plotone{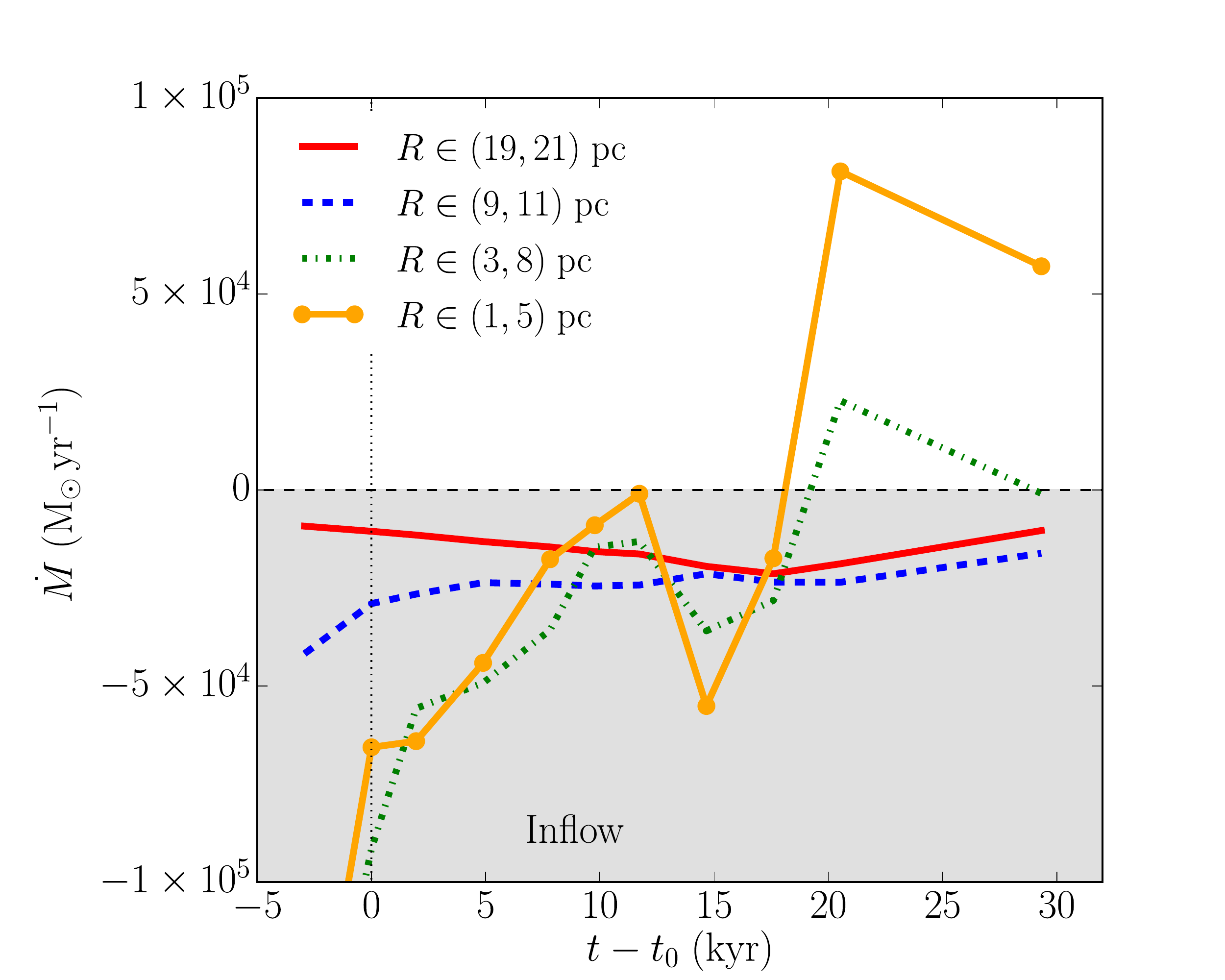}
\figcaption{Time evolution of the gas accretion rate at different radii from the center of the merger remnant.
The accretion is computed inside cylindrical shells of inner and outer radii marked in the legend and vertical thickness 2 pc (see the text for further details).
\label{fig_accretion}}
\end{figure}

Figure \ref{fig_accretion} shows the time evolution of the gas mass inflow, which we interpret as a gas accretion flow onto the central disky core.
The mass inflow rate  is measured inside cylindrical shells with vertical thickness of 2 pc as 
$\dot{M} = \Delta R^{-1} \sum_{j} m_{j} V_{R,j}$, where $\Delta R$ is the radial width of the shell, $m_{j}$ and $V_{R,j}$ are the mass and the radial velocity, respectively, of the $j$-th gas particle inside the annulus.
Figure \ref{fig_accretion} shows that accretion rates of $\sim 10^{4}$ $M_{\odot}$~yr$^{-1}$ are sustained from $\sim 20$ pc, where the radial motions dominate the dynamics of the gas, down to the scale of the inner compact disc\footnote{The outflow occurring at the final stage of the evolution in the inner region are due to the presence of a second massive core; see Section \ref{sec_evol} for details.} $\sim 1$ pc, where the gas is rotationally supported (see Figure \ref{fig_velocity}).

\begin{figure}
\epsscale{1.2}
\plotone{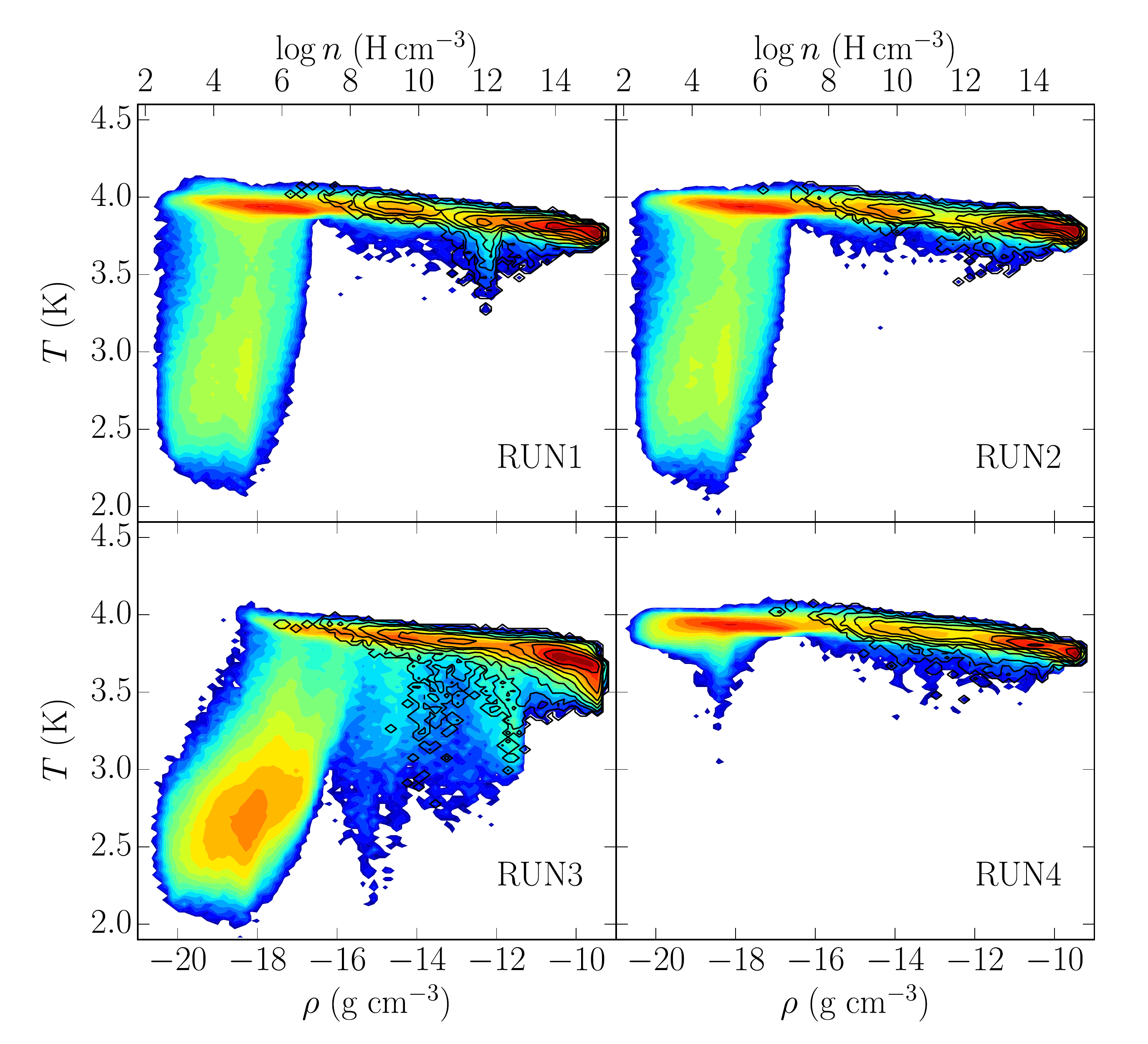}
\figcaption{Phase diagrams of the particles within a 50 pc volume for the same runs shown in Figure 1, at $t=t_0 + 5$ kyr.
From top-left to bottom-right, we show results for RUN1, RUN2, RUN3 and RUN4.
Black contours are superimposed to show the particles that are located in the inner parsec region, within the inner compact disk (see text).
\label{fig_PhaseDiag}}
\end{figure}

The interplay between the strong shock heating associated with the highly  supersonic inflow and radiative cooling is such that the nuclear disk maintains 
a temperature $\sim 3000-6000$ K for $t > t_{0}$.
This is shown in Figure \ref{fig_PhaseDiag}, where we compare the phase diagrams of all the runs at $t=t_0 + 5$ kyr, highlighting the nuclear disk
region with black contours.
The phase diagrams are very similar in the high density region occupied by the nuclear disk, regardless of the particular cooling implementation (i.e. ``metal cooling'' with or without the ``thermal balance model'') and the specific formulation of the SPH hydrodynamic force.
In the absence of any cooling, an adiabatic, strong shock at the infall velocity found in the simulations ($\sim 1000$ km~s$^{-1}$, see the text above and Figure \ref{fig_mach}) would induce a post-shock temperature $\gtrsim 10^6$ K.
However, the presence of cooling regulates the temperature in the whole nuclear disk ($\lesssim 5$ pc) to a few thousand K 
(note that the cooling rate drops significantly below $10^4$ K even for solar metallicity gas).
The similarity between the various runs suggests that shock heating during supersonic infall, which is equally present in all of them, 
plays a major role in balancing cooling and setting the temperature of the core.
Likewise, the diagrams show  that the ``metal cooling'' is more effective at lowering the temperature in the low density, low Mach number regions far from the core.
The latter appears less effective when GDPSH is employed (RUN4), likely because the new SPH implementation facilitates mixing between different gas phases when used in conjunction 
with diffusion (Wadsley et al., in preparation).
Finally, cooling is enhanced at both low and intermediate densities when the thermal balance model is used (RUN3) because in such a model  the cooling rate for solar metallicity
gas is increased at low and intermediate densities, before absorption and re-radiation by dust suppresses cooling at high optical depths, as shown in Roskar et al. (2014).   
We conclude that {\it the warm temperature maintained by the central disk-like core seems to be a robust product of the supersonic gravitational infall established by 
the larger scale dynamics of a major galaxy merger such as the one simulated here}.
We also note that no star formation occurs in the warm core simply because the gas never cools enough to
meet the temperature condition of our star formation prescription (see Section \ref{sec_suite}).
Instead, sporadic star formation can occur in the densest gas pockets further away from the center, where metal-line cooling is effective.

Since the the structure and thermodynamics of the inner region within $\sim 5$ pc does not change appreciably among the different runs, in the following we will focus on the results of RUN4, which employs both GDSPH and diffusion.
This simulation should capture better mixing as well as the multi-phase structure of the flow \citep{keller+14}.


\subsection{A supermassive compact disk in a gravoturbulent, optically thick inflow}\label{sec_stability}

\begin{figure*}
\plotone{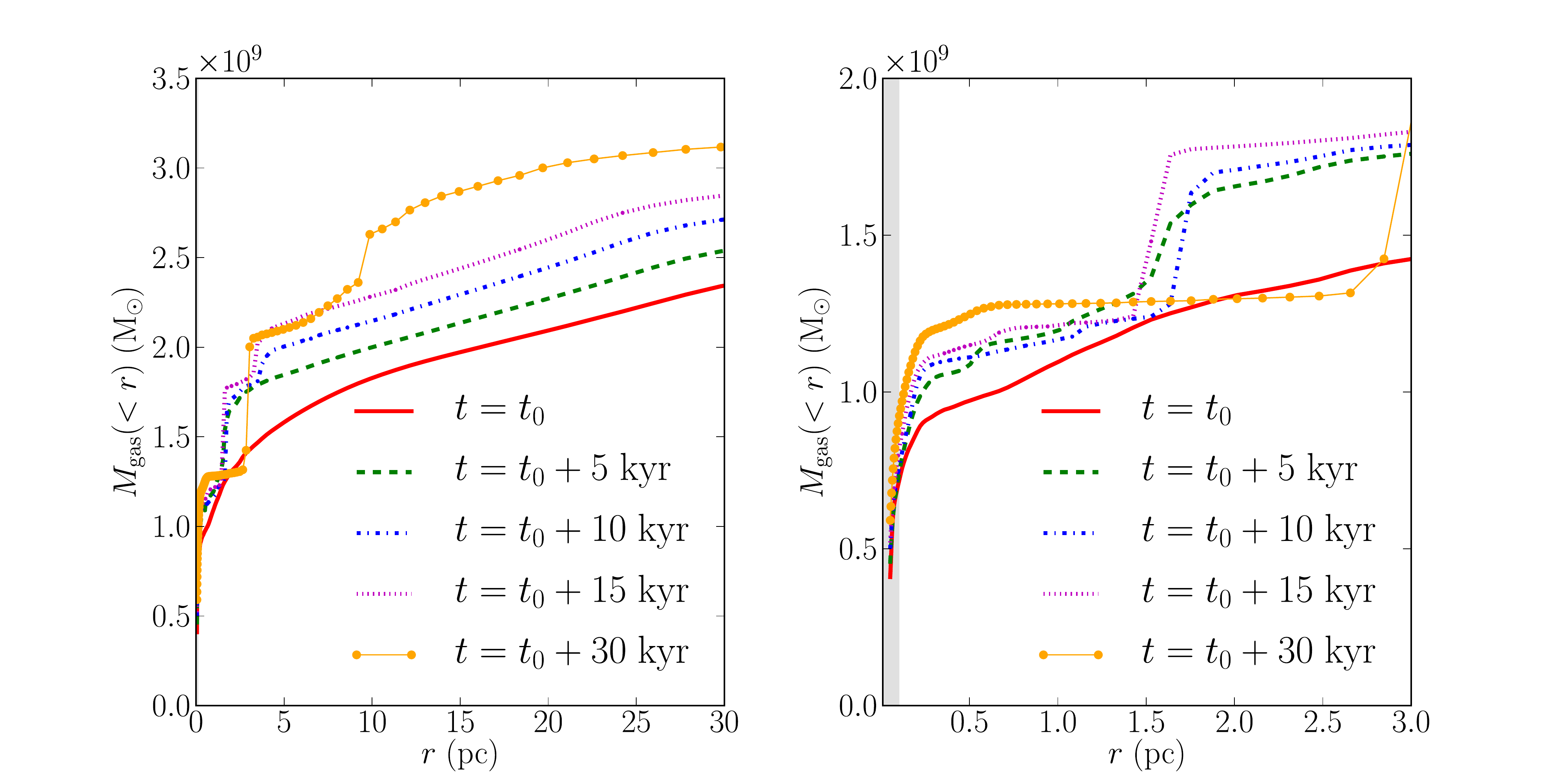}
\figcaption{Time evolution of the enclosed gas mass inside 30 pc (left) and inside 3 pc (right), for RUN4.
The reference time $t_{0}$ is the time of the coalescence of the two merging cores (see text for details).
The gray stripe in the right panel highlights the resolution limit set by the gravitational softening.
\label{fig_mass}}
\end{figure*}

Here we discuss in detail the structural properties and evolution of the central nuclear disk.
Figure \ref{fig_mass} shows the cumulative mass measured inside cylindrical shells with vertical thickness of 3 pc.
Immediately after formation, the compact disk-like core forms and weighs  $\lesssim 10^9$ $M_{\odot}$, with variations of less than a factor of 2 between the different runs, and is barely $0.5$ pc in size (see Figure \ref{fig_SD}).
It has has a mean temperature of $\sim 4000-5000$ K and an aspect ratio $\sim 0.05-0.1$.
It is born with marked non-axisymmetric structure due to its high self-gravity and mass accretion rates $\sim 10^{5}$ $M_{\odot}$~yr$^{-1}$, which are almost an order of magnitude higher than those found in MA10 with an effective EoS.
Then, the inner compact disk grows inside-out as it accretes infalling gas with increasingly higher angular momentum, reaching a mass of of $1.3 \times 10^{9}$ $M_{\odot}$ within $\lesssim 1$ pc and forming a nuclear disk with an extension of $\lesssim 5$ pc after 30 kyr.
Such  a disk size is a factor of 10 smaller than the radius reached by the disk in MA10 soon after the merger is completed.
The disk is much denser than in MA10 because the infalling gas here is much colder than with an effective EoS.
The cold infall also enhances the free-fall velocity $v_{\rm ff}$ since the baryonic mass density increases more and faster relative to MA10, an effect analogous to inside-out protostellar collapse \citep[see e.g.][]{stahler+05}.
This in turn increases the mass inflow rate since $\dot M  \sim {v_{\rm ff}}^3/G$ (at least until the effect of angular momentum becomes important). 
The free-fall time is also proportionally reduced as $t_{\rm ff} \sim R_{\rm cl}/v_{\rm ff}$, where $R_{\rm cl}$ is the size of the collapsing region. 
Not surprisingly, the inner compact disk is established with a central density exceeding $\sim 10^{-10}$ g~cm$^{-2}$, comparable to the density that in MA10 was reached only after $100$ kyr as a result of a secondary infall inside the disk triggered by spiral instabilities.
{\it Therefore, owing to radiative cooling the dynamical evolution of the nuclear region is both stronger and accelerated in these new simulations relative to MA10}.

The fast accretion rate on to the nuclear disk/inner core continuously maintains a high level of self-gravity (see e.g. \citealt{boley+09} and \citealt{hayfield+11} for an analogous behavior in protostellar disks).
This induces the nuclear disk to develop global spiral modes at a few parsec scale and an oval distortion within the central parsec (see Figure \ref{fig_SD}).
In particular, we expect that a further increase in spatial resolution would turn 
the oval distortion into a strong bar-like mode since we measure $T_{\rm rot}/W \sim 0.1$ within 0.5 pc, where $T_{\rm rot}$ is the rotational kinetic energy and $W$ the gravitational binding energy.
This value is indeed close to the condition $T_{\rm rot}/W > 0.1-0.3$ for bar formation in 3D rotationally flattened, self-gravitating axisymmetric 
configurations \citep{durisen+85,pickett+96, christodoulou+95}, where the exact threshold value depends on the equilibrium structure  of the configuration \citep{christodoulou+95}.
The value of $T_{\rm rot} / W$ that we measure is likely reduced by gravitational softening, which damps the growth of non-axisymmetric instabilities,
a well documented fact in galactic disk-scale simulations \citep{mayer+04,debattista+06,kaufmann+07}.

We expect that the unresolved bar mode in the core will sustain an inflow down to $\ll 10^{-1}$ pc, as found in 
circumnuclear disk simulations of \citet{choi+13}.
This is also in line with the general picture of bars-in-bars proposed by \citet{begelman+06} (see also \citealt{begelman+09}), 
except for the much higher inflow rates relative to isolated protogalactic disks owing to the trigger from larger scales provided by the  merger dynamics.
As the system will lower its rotational energy due to angular momentum transport the bar mode mode will weaken, but other
instabilities can intervene at lower $T_{\rm rot}/W$, for example one-armed modes \citep{pickett+96}.
Considering the conservative case of one-armed modes, less effective than bars at shedding angular momentum, a decrease of a factor of 2 
in the specific angular momentum is still expected after a few core rotations \citep{pickett+96}. 

\begin{figure}
\epsscale{1.1}
\plotone{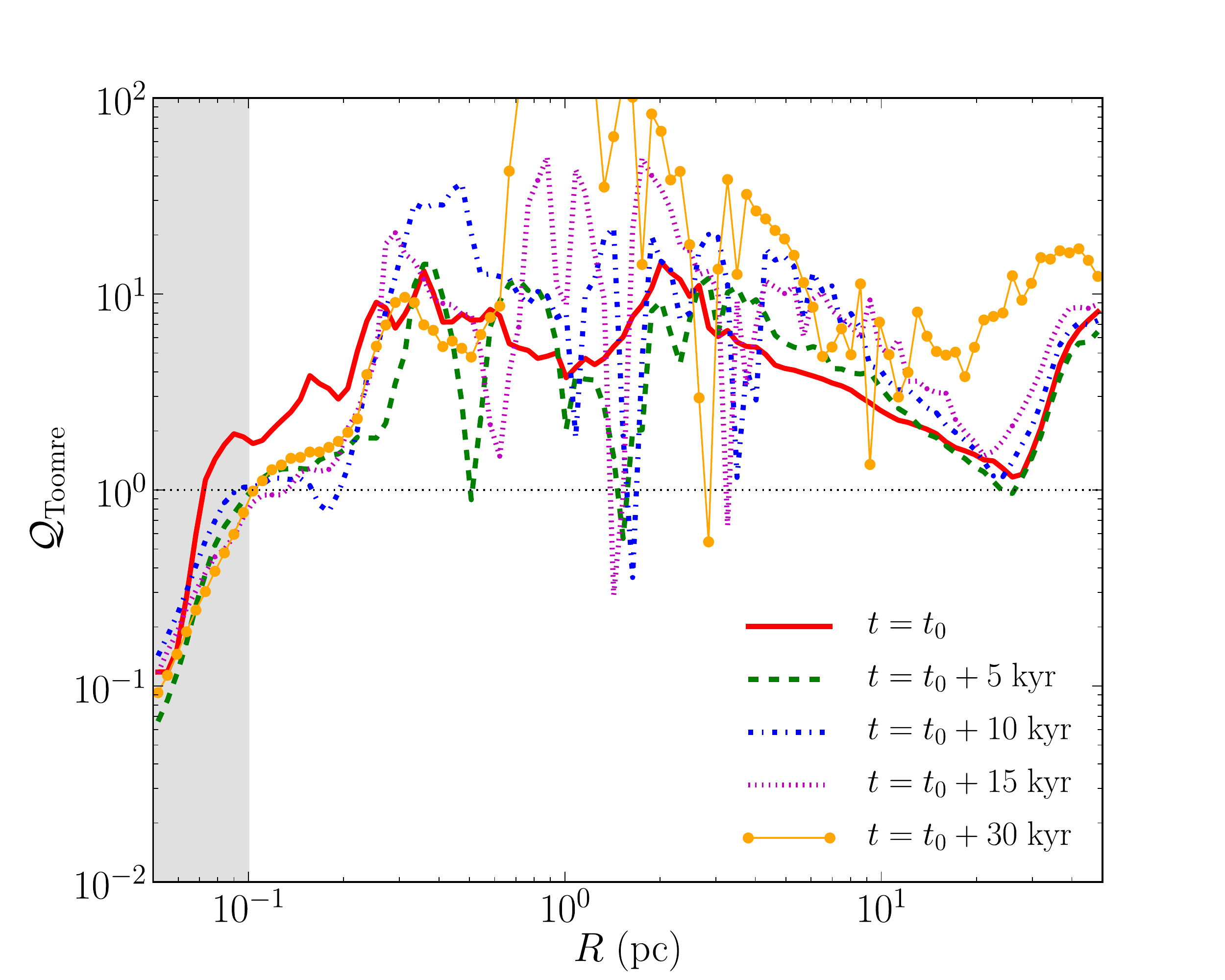}
\figcaption{Time evolution of the Toomre parameter profile measured from RUN4 (see the adopted definition of $\mathcal{Q}_{\rm Toomre}$ in the text).
Dips correspond to regions where clumps form from gas fragmentation along spiral arms, predominantly outside the inner pc region (see Figure \ref{fig_SD} for comparison on the spatial distribution of clumps).
The dotted horizontal line marks $\mathcal{Q}_{\rm Toomre} = 1$ for reference.
The gray band highlights the resolution limit given by the gravitational softening.
\label{fig_Q}}
\end{figure}

We now examine the conditions of the self-gravitating gas flow in the central disk more closely.
Fragmentation in a (laminar) rotating gas disk requires two conditions: (i) the Toomre parameter\footnote{The conventional stability threshold $\mathcal{Q}_{\rm Toomre} > 1$ is derived using linear perturbation theory for an infinitesimally thin, axisymmetric disk. However, higher order perturbation theory and numerical simulations in 2D and 3D favor the 
phenomenological threshold $\mathcal{Q}_{\rm Toomre} > 1.5$ in more general conditions \citep[e.g.][]{durisen+07}.} 
$\mathcal{Q}_{\rm Toomre} < 1.5$; \emph{and} (ii) the local cooling timescale\footnote{The cooling timescale condition is well established in simulations of self-gravitating gas disks \citep[e.g. self-gravitating proto-planetary disks;]
[]{gammie+01,rice+03,mayer+05,meru+11a,helled+13}.
However, the exact value of the ratio between the cooling time and the orbital time is still a matter of debate since proof of convergence as a function of resolution and cooling law implementation has not been achieved yet \citep[see e.g.][]{meru+11b,helled+13}.
Therefore, we will be intentionally conservative and assume  that a cooling time equal to the orbital time is a sufficient second condition for fragmentation.}
has to be shorter than the local orbital time \citep[e.g.][]{lodato+06,durisen+07}.
Figure \ref{fig_Q} shows the time evolution of the $\mathcal{Q}_{\rm Toomre}$ profiles of our simulation RUN4.
We compute the Toomre parameter taking into account supersonic turbulence in the gas as $\mathcal{Q}_{\rm Toomre} = v_{\rm turb} \kappa / (\pi G \Sigma)$, where $\kappa$ is the epicyclic frequency, $\Sigma$ is the gas surface density, and $v_{\rm turb} = \sqrt {\sigma_{\rm g}^2 + c_{\rm s}^2}$. 
Here, $\sigma_{\rm g}$ is the 3D velocity dispersion of the gas (measured as in Section \ref{sec_global}) and $c_{\rm s}$ the thermal sound speed.
The $\mathcal{Q}_{\rm Toomre}$ stability threshold is satisfied in most of the  nuclear disk after a few thousand year since its emergence, 
except for some minima below 1.5 typically arising at $1 \lesssim R/{\rm pc} \lesssim 5$ along spiral arms in the nuclear disk (see Figure \ref{fig_SD}),
which indeed is where gaseous clumps are observed to form via fragmentation.
Supersonic turbulence provides support against gravitational collapse, both directly (kinematically) and via heating through small scale shocks as typically observed in star formation simulations \citep{padoan+97}.
Therefore, turbulence influences also the cooling time criterion increasing the effective cooling time.
Although we can not probe directly the non-linear coupling between turbulence heating and radiative cooling, we argue that supersonic turbulence plays 
an important role in regulating the temperature of the nuclear region, since $\sigma_{\rm g} \gg c_{\rm s}$ and the runs appear rather 
insensitive to the details of the cooling within a few parsecs.
The key role of turbulence in regulating disk fragmentation was noted also by \citet{choi+13} and \citet{latif+13}.

We do not include radiative transfer in our simulations, although the ``thermal balance model''  can be viewed as an approximate way to 
introduce radiative transfer effects. Nevertheless, in order to support further our results we
can show analytically that the intrinsic optically-thick nature of the nuclear disk would suppress fragmentation in the inner parsec.
Indeed, the typical optical depth in the disky core within 1 pc is $\tau_{\rm es} \sim N_{\rm g} \sigma_{\rm T} \sim 10^4$, where $\sigma_{\rm T} \simeq 6.65 \times 10^{-25}$ cm$^{2}$ is the Thomson scattering cross section and $N_{\rm g} \sim 10^{29}$ H~cm$^{-2}$ is the mean gas column density within $0.5$ pc (corresponding to a surface density $\sim 10^{9}$ $M_{\odot}$~pc$^{-2}$).
Note that $N_{\rm g}$ varies by an order of magnitude above an below the quoted value in the very inner region and in the low density gaps between spiral arms and rings, respectively.
At the temperature of the core, which is $\sim 3000-5000$ K, the opacity due to electron capture by H$^{-}$ might indeed be up to ten 10 times higher than $\sigma_{\rm T}$ for the densities in the inner parsec ($\sim 10^{-10}$ g~cm$^{-3}$), making our estimate of the optical depth conservative.
Finally, the gas would then cool on the photon diffusion timescale $t_{\rm diff} \sim H \tau_{\rm es} / c \sim 3000$ yr, where $c$ is the speed of light and $H \sim 0.1 R \sim 0.1$ pc is the vertical scale height of the disk.
This timescale is much longer than the orbital time, which is always $< 500$ yr for $R < 1$ pc, meaning that no fragmentation should occur.
Note for comparison that \citet{ferrara+13b} determined a similar diffusion timescale\footnote{They assumed a lower optical depth but a higher value for $H$.}, 
but they compared it with a free-fall time two orders of magnitude longer then the orbital time we find in our simulation, concluding that fragmentation would be ubiquitous in the post-merger core.
This is because they underestimated the central density of the disk due to their isochoric assumption.
In other words, they explored the effect of radiative cooling without considering how both
cooling and shocks in the infalling gas would modify the density structure of the disk in the first place.

The conditions in the disk outside $\sim 1-2$ pc change because the average velocity dispersion decreases, the orbital time increases and fragmentation can take place
(see upper panel of Figure \ref{fig_SD}).
The co-existence of an inner region stable to fragmentation and an outer region unstable to fragmentation is reminiscent of massive self-gravitating protostellar and protoplanetary disks \citep[e.g.][]{boley+10,helled+13}.
Large scale spiral instabilities are present, and massive clumps in the range $10^6-10^8$ $M_{\odot}$ form along the arms.
Most of them appear to be only marginally bound, being dispersed and reforming on a few orbital timescales.
The typical temperature of the clumps is $> 2000$ K, hence no conventional star formation can take place inside them.
Clumps that are dense enough to survive tides will spiral towards the center as a result of dynamical friction on a few orbital times \citep{noguchi+98,bournaud+09,fiacconi+13}.

\begin{figure}
\epsscale{1.1}
\plotone{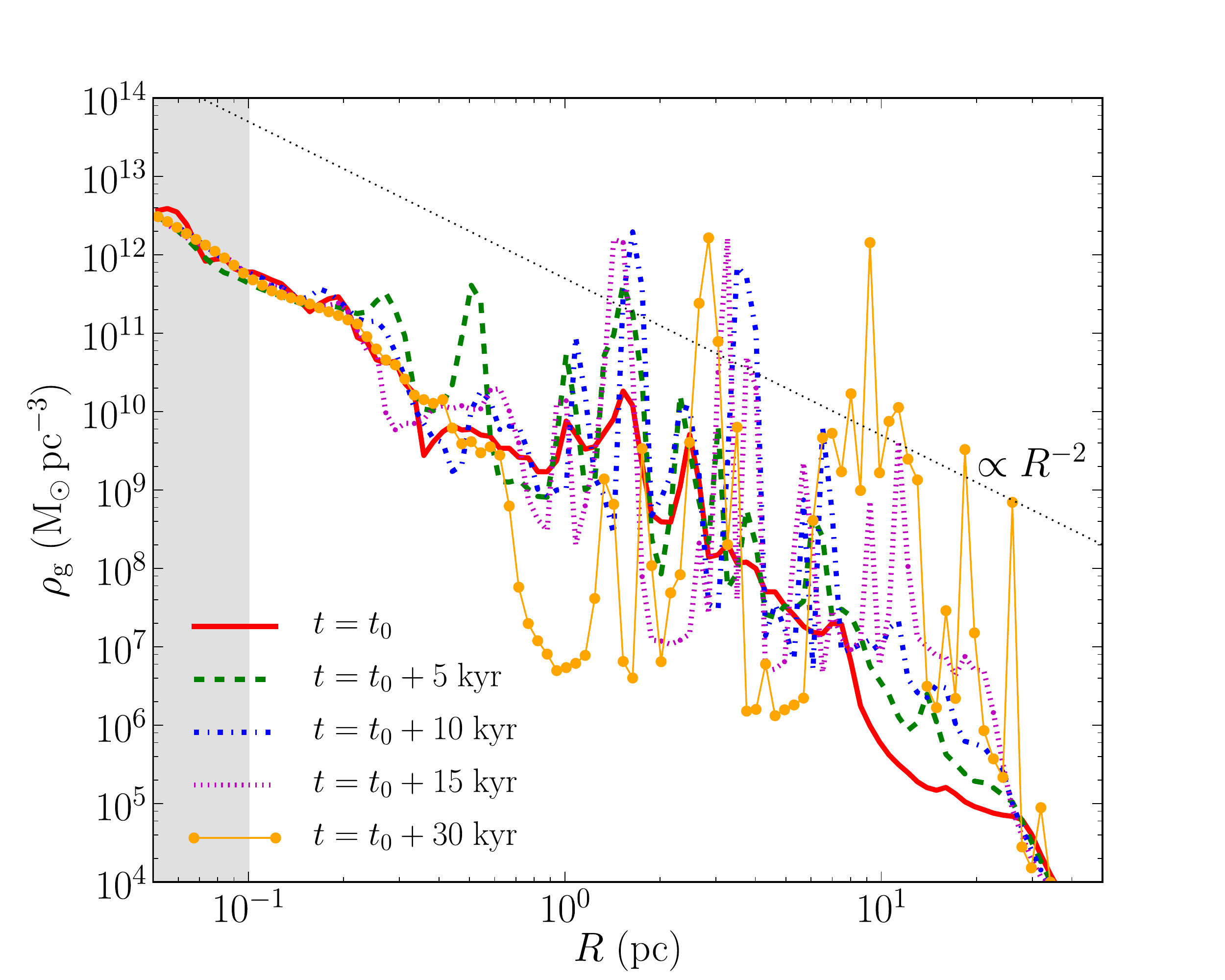}
\plotone{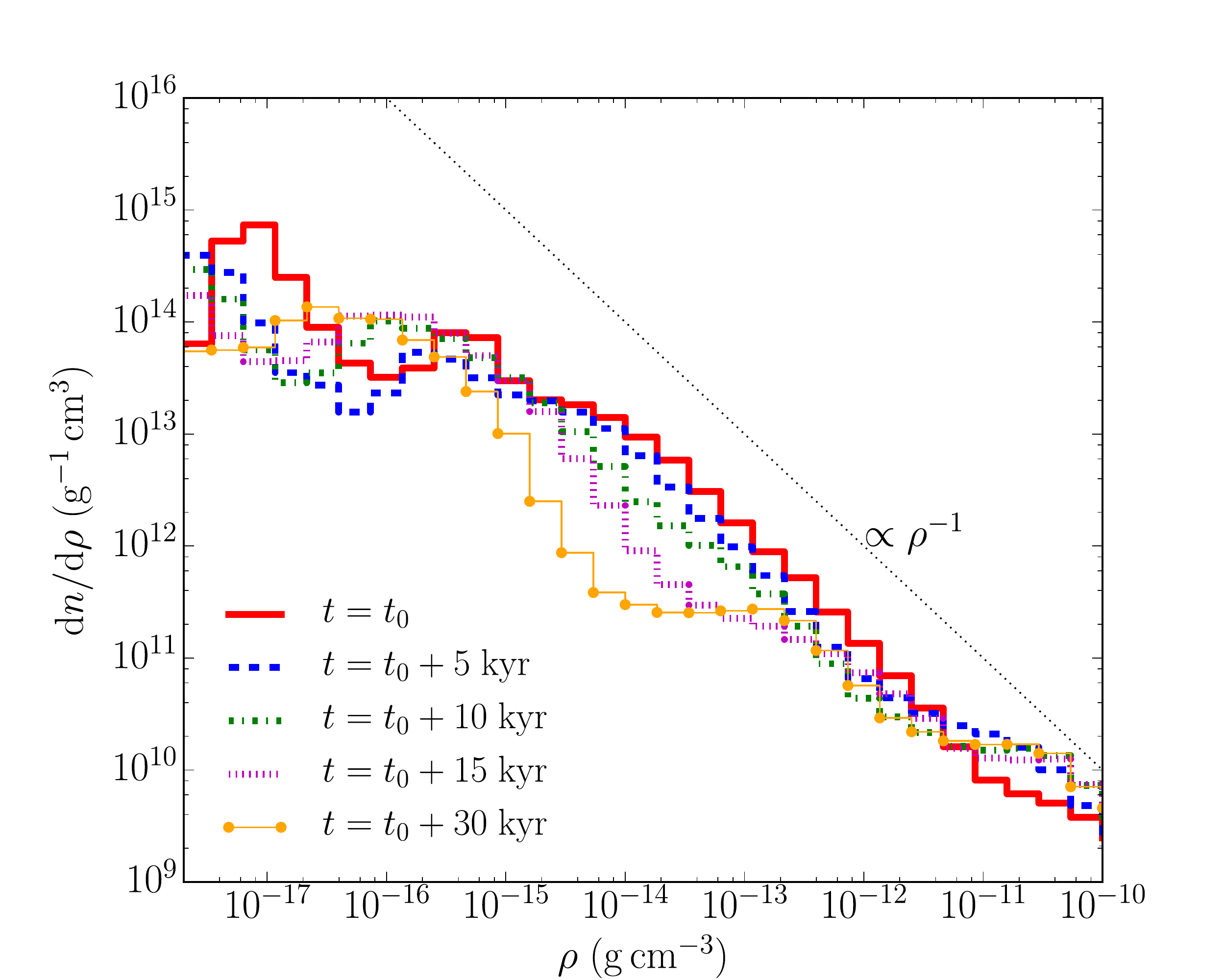}
\figcaption{Time evolution of the gas density radial profile for RUN4 (top).
The dotted line show the $\propto R^{-2}$ scaling.
Time evolution of the gas density PDF (bottom).
The dotted line show the $\propto \rho^{-1}$ scaling.
The density PDF is computed inside the cylindrical region with radius of 3 pc and with height of 2 pc centered on the nuclear disk.
\label{fig_dens}}
\end{figure}

The gravoturbulent nature of the flow is reflected also by the density profile and the density probability distribution function (PDF) acquired by the gas, shown respectively in the upper and lower panel of Figure \ref{fig_dens} at different times.
The mass density profile achieved in the inner 10 pc as a result of the coincident action of self-gravity, turbulence and thermodynamics is close to $\propto R^{-2}$.
Such a density slope has been shown to favor central collapse relative to fragmentation in the envelope for gravoturbulent molecular clouds, a condition that is often invoked to form a massive star rather than a cluster of lower mass stars \citep{girichidis+11}.
This qualitatively corresponds to the behavior observed in our simulations.
The density PDF of the gas in the nuclear region carries the imprint of supersonic turbulence in presence of self-gravity.
It exhibits a slope close to $-1$, in nice agreement with the simulations of \citet{choi+13}, as opposed to the slope of $\sim -2$ typically found before self-gravitating collapse ensues in simulations of supersonic turbulence in the ISM \citep[e.g.][]{scalo+98,kritsuk+11}.
It has been suggested that a slope $\sim -1$ may result from dynamically important angular momentum \citep{kritsuk+11}, and indeed we verified that such region corresponds to the region where the nuclear disk assembles, again in agreement with \citet{choi+13}.
However, our PDF cannot be simply described by a log-normal distribution function at low densities as in \citet{choi+13}.
We argue that the excess amplitude relative to a log-normal PDF at densities $\rho < 10^{-15}$ g~cm$^{-3}$ is due to the large amount of infalling gas 
exterior to the inner compact disky-core.
Presumably less gas is present at large radii and lower density in \citet{choi+13} since these authors model an isolated protogalaxy rather than a galaxy merger.
Our interpretation is supported by the fact that the low density tail tends to become slightly lower with time as less infalling continues 
to reach the inner few parsecs.


\section{Possible evolutionary pathways of the inner compact core}\label{sec_evol}

We stop our simulations between 30 and 50 kyr after $t_{0}$ because we reached prohibitively small timesteps and the central inflow 
is saturated at the resolution scale set by the gravitational softening.
Although we cannot ascertain the ultimate fate of the gas below the resolution limit, the simulations provide a wealth of information that allows us to delineate possible 
scenarios to form a massive BH seed given the final conditions inside the pc-size compact disky core present in all of our runs.
Hereafter we will refer to the final conditions of RUN4 described in the previous section.

The mass accretion rate has been shown to be the key factor deciding whether a SMS will form or the system will remain in a protostar-like phase, on the 
Hayashi track, and produce directly a quasi-star \citep{begelman+10,schleicher+13,choi+13}.
\citet{schleicher+13} found that the critical mass necessary for forming a SMS is $3.6 \times 10^8~\dot{m}^3$ $M_{\odot}$, where $\dot{m}$ is the mass accretion rate normalized such that $\dot{m}=1$ for an accretion rate $\dot{M}=1$ $M_{\odot}$~yr$^{-1}$.
Such critical mass is the mass above which the accretion timescale becomes longer than the Kelvin-Helmoltz timescale in the stellar interior, implying that the protostellar core will contract, the star will leave the Hayashi track and will become a full fledged SMS.
The accretion rates found in all our runs in the inner compact disk are typically $> 10^4$ $M_{\odot}$~yr$^{-1}$.
They correspond to $\dot{m} > 10^4$ and to a critical mass to form a SMS $> 10^{12}$ $M_{\odot}$.
This mass is higher than the entire mass reservoir contained within several hundred parsec, suggesting that a SMS would not form. Instead, a QS could form
in $\sim 10^7$ yr, after the central region has collapsed into a small BH \citep{schleicher+13}.
We caution that we cannot determine the actual radius of the protostar from of our simulations, which should be smaller than our gravitational softening 
\citep{hosokawa+12, hosokawa+13}, hence we cannot exclude a reduction of the accretion rate at the scale of the protostar.
Moreover, we recall that \citet{schleicher+13} derived their results under the assumption of a spherical, adiabatic, steady contraction without rotation as opposed to a highly dynamical, angular momentum loaded flow as that in our simulations.

Conversely, another condition may become relevant on much shorter timescales if the
high inflow rates measured in our simulations can be sustained at scales below our resolution.
This is the condition for global relativistic radial collapse in the strong field regime.
Numerical general-relativistic (GR) simulations show that the instability for a rotating fluid configuration (a polytrope with $\gamma = 4/3$) is reached for: (i) a \emph{compactness threshold} $R < 640 \, GM/c^2$, and (ii) a dimensionless spin parameter $q \equiv cJ/(GM^2) < 0.97$ \citep{baumgarte+99,shibata+02,saijo+09,reisswig+13}, where $M$ is the mass and $J$ the total angular momentum of the cloud.
This has been shown to apply to both uniformly rotating and differentially rotating clouds. 
The inner disky core in our simulations reaches quickly $\sim 10^9$ $M_{\odot}$ within $0.2$ pc and it has typically $q \sim 15$.
On the other hand, the compactness threshold for $M \sim 10^9$ $M_{\odot}$ is  $R \sim 0.03$, which is $\sim 6$ times smaller than the characteristic
radius of $0.2$ pc.
Therefore, the disky core would seem to be stable to the general relativistic (GR) radial instability.
However, as we outlined in section 3.2 we expect that it will become bar-unstable, which would transport angular momentum outward and cause further contraction, 
possibly pushing it closer to the verge of the instability.

An alternative, perhaps more effective way to discuss stability is to rely on another result of
relativistic simulations. Indeed, starting from 2D and 3D rotating
polytropic clouds with masses exceeding $10^6$ $M_{\odot}$,  these simulations show that $T_{\rm rot}/W \sim 0.01$ or lower is 
a sufficient condition to bring the cloud to the radial collapse stage under a small initial perturbation
\footnote{Recently \citet{montero+12} have shown that if nuclear burning via the CNO cycle begins at the very center it can induce a shock wave
that deflagrates the protostar but so far this has been obtained only in 2D simulations, and for systems with mass $10^5-10^6$ $M_{\odot}$.}
\citep{shibata+02}.
This is of course a phenomenological criterion, and may change with a different EoS, but offers a useful guideline.
If we adopt  the latter condition, the angular momentum in the inner compact disk has to decrease substantially to enter radial collapse 
since $T_{\rm rot}/W \sim 0.1$ at radii $\lesssim 0.5$ pc.
Drawing from the calculations of bar-unstable protostellar clouds, which can apply here since the eventual subsequent contraction will mostly be in the newtonian 
regime, one expects a decrease of the specific angular momentum $j$ by a factor of 2 over a few dynamical times, \citep[e.g.][]{pickett+96}, i.e. over
$< 10^4$ yr.
At the same time, the mass of the system can grow up to a factor of $\sim 2$, if accretion rates $> 10^4$ $M_{\odot}$/yr are sustained by the bar down to scales $< 0.5$ pc for
$\sim 10^5$ yr.
Therefore, $T_{\rm rot}/W$ would decrease by almost an order of magnitude at fixed radius on relatively short timescales (since it scales as $T_{\rm rot}/W \propto 
j^{2} M^{-1} R^{-1}$), reaching the critical threshold for radial collapse.

Although we cannot reach firm quantitative conclusions, our simple estimates suggest the direct massive BH formation via GR radial instability might occur 
if gas inflow rates as those measured un our simulations are maintained down to scales close to our resolution limit for a time only slightly longer than
we could probe here. 
Since up to $\sim 60-90\%$ of the progenitor mass can be retained during GR radial collapse \citep[e.g.][]{saijo+09,reisswig+13}, the emerging 
BH seed would have a mass roughly between $10^{8}$ and $10^9$ $M_{\odot}$, namely in the SMBH mass range and, most importantly, already close the mass inferred 
for the SMBHs powering the high-$z$ QSOs.

\begin{figure*}
\epsscale{1.2}
\plotone{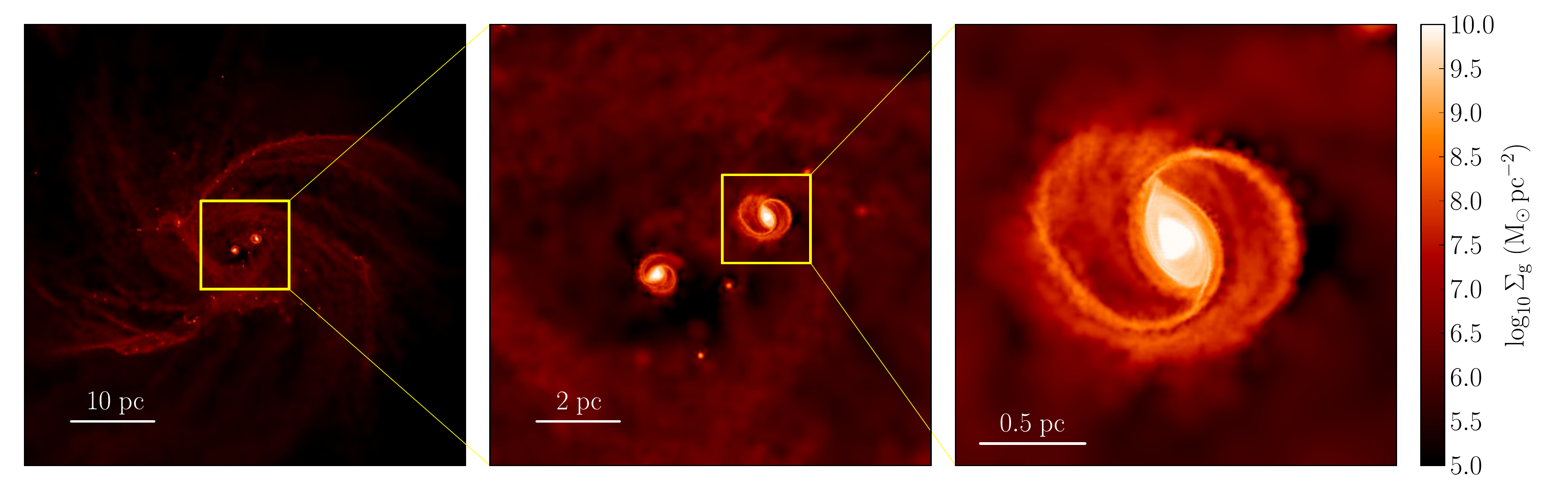}
\figcaption{Projected gas density maps of a snapshot at time $t = t_{0} + 30$ kyr, showing the formation of a secondary massive clump as a result of fragmentation (see text).
Each panel, from left to right, shows a different scale. 
In particular, a zoom-in on the primary core is shown on the right.
\label{fig_final}}
\end{figure*}

Finally, another intriguing aspect of the simulations is that, at least in RUN4, a binary system of two compact disk-like cores eventually arises.
This is shown in the gas density projection in Figure \ref{fig_final} at time $t_0 + 30$ kyr.
The secondary clump forms after about 25 kyr from fragmentation along  a prominent spiral arm.
It has a mass only 5 times smaller than that of the primary, which is $\gtrsim 10^9$ $M_{\odot}$.
Its appearance is also highlighted by the bump at $r \sim 3$ pc in the cumulative mass profile shown in Figure \ref{fig_mass}.
Simulations of massive black hole binaries which evolve in a similar environment suggest that the
separations of the two objects will be reduced to $< 1$ pc by dynamical friction against the background in less than 1 Myr \citep[e.g.][]{mayer+13}.
If a massive BH seed forms in the meantime at the center of both objects one may speculate that a merger between two massive BHs will occur, leading to a burst 
of gravitational waves (GWs).
The frequency of such signal may vary a lot depending on the mass ratio of the BH seeds and on their orbital eccentricity.
Assuming conservatively the BH seed mass found in models of direct collapse that posit a quasi-star stage
($\sim 10^3-10^4$ $M_{\odot}$; \citealt{dotan+11}), the signal could be detected by the {\it eLISA} probe \citep{amaro-seoane+13}.
Alternatively, if a secondary massive clump is not formed during the stage studied here, the central core might still fragment later during the relativistic 
phase of the collapse.
Indeed in the latter phase a relativistic bar instability could be triggered, which may lead to fission into two or more massive BHs. 
This was demonstrated by \citet{reisswig+13} with 3D relativistic hydrodynamics simulations starting 
from polytropic, marginally unstable, differentially rotating SMS.

We caution that the arguments outlined in this section do not address the thermodynamical evolution during the
final collapse stage, which will not necessarily produce a simple polytropic configuration.
Hence, only by evolving the end state of our system with a numerical GR code along with a proper treatment of
radiation hydrodynamics we will be able to ascertain whether or not direct formation of a SMBH is possible.


\section{Discussion and Caveats} \label{sec_4}

We have presented simulations with greater physical realism than those in our previous work (MA10).
Our results lend stronger support to the scenario presented in MA10 and \citet{bonoli+14}, in which the SMBHs powering the bright QSOs
at $z > 5$ form by direct collapse driven by multi-scale gas inflows in major mergers between the most massive galaxies already present at $z \sim 8-10$.
A key difference with all other direct collapse models presented in the literature is that here we do not require metal-poor gas or dissociation of H$_2$ to 
suppress cooling, rather radiative cooling itself fosters the rapid formation of a compact and dense nuclear disk.
Fragmentation can happen, but the inner pc-scale core is stabilized by shock heating, gravoturbulence and the high optical depth of the gas.
The gas collapses faster and it reaches higher densities relative to MA10 simulations, which adopted and effective EoS rather than incorporating cooling.
While the subsequent transport of angular momentum below $0.1$ pc will have to be explored further, we have argued, based on simple analytical arguments, that rapid removal of angular momentum by non-axisymmetric instabilities might continue below the resolution of our simulations.
Most importantly, we have highlighted the possibility that the compact disky-core, which encompasses a mass of nearly $10^9$ $M_{\odot}$,  
may reach the condition for the onset of the GR radial instability by collapsing just an extra decade in radius.

When such a condition is reached, GR simulations tell us that the inner core would leave behind a BH seed exceeding half of its mass.
Being really conservative, we can posit that the BH seed will have a mass of $10^8$ $M_{\odot}$.
With such a seed, little accretion will be needed to reach rapidly the BH masses inferred for high-$z$ QSOs.
Indeed, the BH would take $\tau_{\rm acc} \approx 10^{8} \, f_{\rm edd}^{-1} \, (\log(M_{\rm \bullet, f}/ M_{\rm \bullet, i})/\log(10))$ yr to 
grow from $M_{\rm \bullet,i}$ to $M_{\rm \bullet,f} = 10 M_{\rm \bullet,i}$.
If we assume an Eddington ratio $f_{\rm edd} \approx 0.3$ all the time, the typical value obtained in radiative simulations of gas accretion onto
BH seeds \citep{milosavljevic+09}, we obtain $\tau_{\rm acc} \approx 3 \times 10^{8}$ yr, comfortably shorter than the time elapsed to $z = 6-7$ in
a $\Lambda$-CDM cosmology. 
Note that the Eddington accretion rate would change from $\sim 2$ to $\sim 20$ $M_{\odot}$~yr$^{-1}$ between $\sim 10^{8}$ and $\sim 10^{9}$ $M_{\odot}$.
The gas reservoir in the inner 100 pc of the merger remnant is $\sim 3 \times 10^9$ $M_{\odot}$ and undergoes infall, hence there is plenty of gas to accrete from, a point already emphasized in MA10.

The thermodynamics of the gas will change drastically after a BH seed forms at the center.  
If the seed emerges already with the mass of a SMBH, any energetic feedback from subsequent accretion, although it can limit its further growth 
is of little relevance for the origin of the high-z QSOs. 
Instead, such an early strong feedback mode might have implications for galaxy formation as it could affect gas cooling
in nearby halos (see e.g. \citealt{dijkstra+06,dijkstra+14}) as well as stifle bulge formation in the host galaxy.
Feedback can surely have a more important impact on the growth of lighter seeds in the range $10^3-10^5$ $M_{\odot}$.
However, we have shown that the photon diffusion timescale is longer than the orbital time below parsec scales and thus longer than the free-fall timescale.
This means that radiation generated by the accretion process will be trapped in the very inner region on characteristic timescale of the inflow itself. 
Therefore, the luminosity produced by accretion on to the BH can be integrated over the time interval $\Delta t$ to yield an estimate of the energy deposited in the inner region, say roughly below 1 pc.
 Such integrated energy release is $E_{\rm fb} \sim \epsilon_{\rm fb} \, L_{\bullet} \, \Delta t$, where $\epsilon_{\rm fb}$ parametrizes uncertainties on the coupling between gas and radiation, and $L_{\bullet}$ is the accretion luminosity on to the BH, which we roughly estimate as the Eddington luminosity for a $10^5$ $M_{\odot}$ BH.
Note $\epsilon_{\rm fb}$ is highly uncertain, but we take it $\sim 0.01$ as usually done in effective models of BH feedback (see e.g. \citealt{dimatteo+05,vanwassenhove+14}).
Over a timescale of $\Delta t=10^5$ yr we obtain $E_{\rm fb} \sim 4 \times 10^{53}$ erg.
This should be compared to the binding energy of the gas, which is $E_{\rm b} \approx G M_{\rm disk}^2/R_{\rm disk} \sim 10^{59}$ erg assuming $R_{\rm disk}=1$ pc and $M_{\rm disk} = 10^9$ $M_{\odot}$.
Therefore feedback should not be strong enough to stifle accretion even if the BH seed accretes up to five orders of magnitude above Eddington. 

Based on the scenario outlined in the previous section, the massive BH formation process takes less than $10^5$ yr
following the merger if GR radial collapse occurs.
This implies that bright QSOs with SMBHs $\sim 10^8 - 10^9$ $M_{\odot}$ can appear essentially as soon as galaxies massive enough to host sufficiently abundant gas reservoirs and multi-scale scale inflows arise.
In MA10 and \citet{bonoli+14} we studied the dependence of the gas inflow on galaxy mass (but fixed gas fraction), finding that a major merger between galaxies having virial mass $> 10^{11}$ $M_{\odot}$ is a necessary condition for the inflow to produce a central collapse.
We can estimate roughly how the inflow rate should scale with virial mass by assuming that the inflow occurs in a self-similar way.
Thus, the inflow rate would scale as $\dot{M} \sim v_{\rm ff}^3/G \propto M_{\rm vir}$, since $v_{\rm ff} \sim V_{\rm circ} \propto M_{\rm vir}^{1/3}$, where $V_{\rm circ}$ is the virial circular velocity of the halo \citep{mo+98}.
In our simulations $\dot M \sim 10^4$ $M_{\odot}$~yr$^{-1}$ and we expect $\dot{M} \sim 10^3$ $M_{\odot}$~yr$^{-1}$ in galaxies ten times lower in mass.
This means that the mass accumulated at the center would be an order of magnitude lower on the same timescale and the inner compact disky-core would not be able to reach the onset of the GR radial instability, unless accretion is sustained at least ten times longer ($> 10^5$ yr).
However, this longer timescale increases the chances that the inflow rate would decrease significantly before reaching the radial instability conditions and 
opens the path to  the formation of an SMS and, eventually, of a QS. 
In this case the scenario would be identical to that adopted in \citet{bonoli+14}, and the seed that would form has a mass in the range 
$10^3-10^5$ $M_{\odot}$ rather than $\gtrsim 10^8$ $M_{\odot}$.
We calculated that about 200 major mergers between halos above $10^{11}$ $M_{\odot}$ take place at $z > 6$ in the {\it Millennium Simulation}, whose volume is $500^3$ $h^{-3} \, {\rm Mpc}^{3}$.
If all these events lead to a SMBH-like seed via GR radial instability with mass $> 10^{8}$ $M_{\odot}$, the number density of quasars would be $\sim 1.6 \times 10^{-6}$ $h^{3} \, {\rm Mpc}^{-3}$.
If we further assume that only halos above $10^{12}$ $M_{\odot}$ produce seeds $> 10^8$ $M_{\odot}$ because of the scaling argument for $\dot{M}$ outlined above, then the number 
of events drops to only about 10, corresponding to  a number density $\sim 3 \times 10^{-8}$ Mpc$^{-3}$
Interestingly, even in this latter case the predicted number density is still high enough to explain all the $z > 6$ QSOs, which have an 
estimated number density of $\sim 10^{-8}$ Mpc$^{-3}$ \citep{willott+10, treister+13}.
We note however that this discussion is based on the scaling arguments derived above, which do not necessarily apply to major mergers.
Therefore, the dependence on galaxy mass  will have to be revisited exploring a wider range of conditions for galaxy mergers, possibly drawn from cosmological simulations.

Some shortcomings are also present in the simulations themselves.
A major limitation is the nature of the adopted initial conditions.
They are still based on binary mergers rather than on mergers drawn from cosmological simulations.
However, a vast body of literature has recently shown that the most massive high-$z$ galaxies have turbulent, clumpy disks (e.g. \citealt{tacconi+10,ceverino+12,forster+14}; but see also \citealt{fiacconi+14} for lower mass galaxies).
These conditions might favor prominent gas inflows during the mergers since non-axisymmetric torquing of the gas will be aided by gas turbulence.
This could lead to higher central gas densities in the galaxy cores prior to merging.
Clumps formed at galactic disk scales would rain down to the center as the merger comes to completion due to dynamical friction, ultimately aiding the build up of central self-gravitating mass.
However, since angular momentum dissipation is crucial, changing the orbit, geometry and properties of the flow by merging misaligned galaxies or galaxies with a more turbulent ISM may change some of the detailed properties of the inflow.
This in turn may affect the properties of the  compact disk-like core emerging at the center of the remnant.
However, it is reassuring that recent studies of the effect of merger geometry on gas inflows down to $\sim 10$ pc find very weak variations \citep{capelo+14}.

Another limitation is that the simulations start from merging galaxies that had no star formation happening at scales below 100 pc because of the initial lack of resolution (i.e. before particle splitting is applied). 
Pre-existing star formation could have stirred the gas in the galaxy cores via stellar feedback prior to the merger, leading to a more fluffy gas distribution at the center of the remnant.
We made an attempt to explore this effect by imposing instantaneous feedback in the central 100 pc region after particle splitting is applied.
To achieve this we force all star particle located within this volume, each of them sampling a Kroupa IMF, 
to have an age of about 6 Myr, so that massive stars readily go off as SN Type II immediately nearly everywhere after the beginning of the simulation. 
This was done in RUN4 and the the resulting effect on gasdynamics was found to be minor. 
The gas closest to the sites of the explosions is heated and remains hot for the whole duration of the simulation\footnote{The blastwave feedback scheme by construction enforces shut-off of radiative cooling for a timescale of $5-20$ Myr \citep{stinson+06}.}, but most of the gas is not directly affected and keeps cooling and flowing inward supersonically.

A potentially more important shortcoming is the lack of feedback provided by the accretion luminosity of the central clump (see \citealt{schleicher+13}), which is the analog of protostellar accretion luminosity in star formation \citep{krumholz+06}.
Drawing from the example of massive stars, radiation pressure associated with accretion luminosity may eventually stop the inflow and generate outflows. 
The largest accretion luminosity is generated in the inner region of the nuclear disk, where the potential well deepens the most.
At $\sim 1$ pc the core generates $L_{\rm acc} = G M_{\rm disk} \dot{M} /R_{\rm disk} \sim 3 \times 10^{46}$ erg~s$^{-1}$ (using
$R_{\rm disk}=1$ pc, $M_{\rm disk} = 10^9$ $M_{\odot}$, $\dot{M} = 10^4$ $M_{\odot}$~yr$^{-1}$; all these quantities evolve during the simulations but by less than an order of magnitude). 
This is smaller  than the Eddington luminosity, $L_{\rm edd} \sim 1.3 \times 10^{47}$ erg~s$^{-1}$, for an object with mass $M_{\rm disk}$, suggesting that radiation pressure should not stop the inflow. 
Nonetheless, during the contraction required in order to reach the stage of the GR radial instability, $L_{\rm acc}$ might increase and become comparable 
to $L_{\rm edd}$.
At the same, the gas optical depth will increase further and might lead to super-Eddington, radiatively inefficient accretion as mentioned above \citep{madau+14}.
Therefore, the dynamical effect of the accretion radiation has to be probed further in order to understand its impact.

The environment in the nuclear disk is highly optically thick, hence detection of the initial stages of this SMBH formation process in the electromagnetic domain may be difficult. 
At a temperature of a few thousand K, the inner disk will shine blackbody photons at visible
wavelengths into the dense infalling gas, which will re-radiate in the infrared,  resembling a protostellar envelope.
The outgoing luminosity might be much less than the accretion luminosity estimated above due to absorption by the larger
scale gas-rich, dusty nuclear environment.
While the {\it James Webb Space Telescope} might be a good candidate instrument to detect  the infrared signal, extinction and 
possibly re-emission to even longer wavelengths by the envelope of the nuclear disk  might make it more  accessible by the
{\it Atacama Large Millimeter Array}.
The detection of a powerful nuclear FIR source  associated with highly supersonic gas infall velocities revealed with high 
resolution spectroscopy could represent the ``smoking gun'' signature of a direct collapse event. 
Yet the spectroscopic accuracy required to detect strong velocity gradients at scales of hundreds of parsecs at $z > 4-5$ 
is hardly attainable with current or upcoming instruments.

Ultimately GWs will provide the cleanest signature of the flavour of direct collapse that we are proposing, namely a ``cold
direct collapse'' into a SMBH driven by the radial instability.
Strong GW bursts with a characteristic wave-form and amplitude do indeed arise during the radial collapse phase 
of (axisymmetric) supermassive objects in numerical GR simulations, having a frequency $10^{-4} - 10^{-1}$ Hz  
that falls within the expected {\it eLISA} band \citep{saijo+09}. 
Furthermore, if binaries of SMBHs arise commonly ``at birth'' in this scenario, as our simulations suggest, 
one would detect two relatively low frequency 
bursts of GWs as the two SMBHs form via the relativistic collapse, followed by an even lower frequency signal 
as the two holes spiral-in towards coalescence a few Myr later \citep{mayer+13}.
A detailed study of the expected  overall pattern and features of the composite signal, 
involving considerations of the timescale  that separates birth and coalescence of 
the two SMBHs, with the aim at assessing detectability by the planned {\it eLISA} interferometer, warrants investigation 
in future  work.

\begin{acknowledgements}
We are grateful for stimulating and helpful discussions with Piero Madau, Monica Colpi, Mitch Begelman, Francesco Haardt, Ralph Pudritz, Chris McKee, 
Mike Fall, Elena M. Rossi, Melvyn Davies, Milos Milosavljevic, Kevin Schawinski, Luciano Rezzolla and Stephan 
Reisswig.  LM, DF and SB acknowledge hospitality to KITP, where crucial phases of this work were conceived and carried out in two different KITP programs in 2013 and 2014 (``A Universe of Black Holes'' and ``Feedback: Gravity's Loyal Opposition'').
\end{acknowledgements}

\end{document}